\documentclass[10pt,english,prb,amsmath,amssymb,showkeys,superscriptaddress,twocolumn,showpacs,floatfix]{revtex4-1}
\usepackage{graphicx}
\usepackage{dcolumn}
\usepackage[colorlinks=true,linkcolor=blue ,citecolor=blue,urlcolor=blue]{hyperref}
\usepackage{multirow}
\usepackage[usenames,dvipsnames]{xcolor}
\usepackage{soul}
\usepackage{braket}
\usepackage{bm}
\usepackage{nicefrac}
\usepackage{siunitx}
\usepackage{color,transparent}
\usepackage[utf8]{inputenc}
\usepackage{upgreek}

\newcommand{\VEC}[1]{\bm{#1}}

\DeclareSIUnit\ML{ML}
\DeclareSIUnit\MLs{MLs}
\DeclareSIUnit\meVA{meV\angstrom^2}

\begin{document}

\title{Fermi energy determination for advanced smearing techniques} 

\author{Flaviano José dos Santos}
\email{flaviano.dossantos@epfl.ch}
\email{flaviano.dossantos@psi.ch}
\author{Nicola Marzari}
\affiliation{Theory and Simulation of Materials (THEOS), and National Centre for Computational Design and Discovery of Novel Materials (MARVEL), \'Ecole Polytechnique F\'ed\'erale de Lausanne, 1015 Lausanne, Switzerland}
\affiliation{Laboratory for Materials Simulations (LMS), Paul Scherrer Institut, 5232 Villigen PSI, Switzerland}
\date{\today}

\begin{abstract}
Smearing techniques are widely used in first-principles calculations of metallic and magnetic materials, where they improve the accuracy of Brillouin zone sampling and lessen the impact of level-crossing instabilities.
Smearing introduces a fictitious electronic temperature that smooths the discontinuities of the integrands;
consequently, a corresponding fictitious entropic term arises, and needs to be considered in the total free energy functional.
Advanced smearing techniques -- such as Methfessel-Paxton and cold smearing -- have been introduced to guarantee that the system's total free energy remains independent of the smearing temperature at least up to the second order.
In doing so, they give rise to non-monotonic occupation functions (and, for Methfessel-Paxton, non-positive definite), which can result in the chemical potential not being uniquely defined.
We explore this shortcoming in detail and introduce a numerical protocol utilizing Newton's minimization method that is able to identify the desired Fermi energy.
We validate the method by calculating the Fermi energy of $\sim$20,000 materials and comparing it with the results of standard bisection approaches.
In passing, we also highlight how traditional approaches, based on Fermi-Dirac or Gaussian smearing, are actually equivalent for all practical purposes, provided the smearing width is appropriately renormalized by a factor $\sim$2.565.
\end{abstract}

\date{\today}

\maketitle

\section{Introduction}


Density-functional theory is a powerful and popular quantum mechanical framework to calculate the ground-state properties of materials~\cite{hohenberg_inhomogeneous_1964,kohn_self-consistent_1965}.
For extended systems studied using periodic-boundary conditions a practical implementation requires integrations in the Brillouin zone, which are typically carried out through discrete sums over finite samplings.
For metals at zero temperature, integrands become discontinuous because the occupation of the electronic states drops abruptly to zero when crossing the Fermi energy.
Without further treatment, extremely fine samplings ($k$-point meshes) are required for accurate calculations.
Moreover, there is no variational principle governing the convergence of the system's total energy with respect to the $k$-point sampling.
This problem is mitigated by smearing approaches~\cite{fu_first-principles_1983,needs_total-energy_1986,methfessel_high-precision_1989,gillan_calculation_1989,de_vita_energetics_1992,de_gironcoli_lattice_1995,marzari_thermal_1999,verstraete_smearing_2001,quantumatk_team_quantumatk_2022}, which introduce a fictitious electronic temperature (or smearing) that broadens and smooths the occupation function, and thus the integrands, leading to much better convergence with respect to the Brillouin-zone sampling.


The most natural choice to add smearing is to introduce an electronic temperature in the physical canonical ensemble~\cite{mermin_thermal_1965}.
However, the slow decaying tail of the Fermi-Dirac function  requires the calculation of a large number of states that would be otherwise unoccupied, thus making computations more expensive; thus, a very popular alternative is to use a Gaussian broadening~\cite{fu_first-principles_1983,needs_total-energy_1986}.
As discussed later, in these smearing approaches the total energy of the system gains an entropic term that is a function of the smearing temperature.
Thus, when typical smearings of a few tenths an eV are used, an a-posteriori correction of the system's total energy~\cite{gillan_calculation_1989,de_vita_ab_1991} is needed to recover the zero-broadening limit, and expensive calculations would be needed to correctly compute the corrections to ionic forces~\cite{de_vita_private_1993, wagner_errors_1998}.
This issue makes relaxations or molecular dynamics simulations impractical since at every step one would need to calculate the derivative of the entropy with respect to positions, which can be obtained from the derivative of the forces with respect to smearing~\cite{de_vita_private_1993}.
Methfessel and Paxton~\cite{methfessel_high-precision_1989} developed a broadening function that yields total free energies independent of the smearing temperature at least up to third order, thus not requiring a-posteriori corrections, and delivering Hellman-Feynman forces consistent with the total free energy.
The Methfessel-Paxton smearing introduces another issue, in the form of a non-monotonic and non-positive-definite occupation function.
As an alternative,  the cold smearing method proposed by Marzari, Vanderbilt, De Vita, and Payne~\cite{marzari_thermal_1999} yields a positive definite occupation function, albeit still non-monotonic, and a free energy independent of smearing temperature up to the second order.
All these approaches allow for a variational or iterative minimization of the free energy functional~\cite{marzari_ensemble_1997}, leading to Hellman-Feynman forces that are the exact total derivatives of the free energy.
This is at variance with other integration schemes, such as the tetrahedron method~\cite{jepson_electronic_1971} and its improved version~\cite{blochl_improved_1994,kawamura_improved_2014}, where one would not have consistency between energy and forces~\cite{kratzer_basics_2019}.

In this paper, we show that the non-monotonic occupation functions of the Methfessel-Paxton and cold smearing broadenings can lead to multiply-defined chemical potentials.
Bisection implementations for the Fermi energy determination can then yield an incorrect solution, especially when applied to semiconductors and insulators, as it is often done in high-throughput approaches where the metallic or insulating nature of the system is a priori unknown.
We suggest instead a novel protocol based on Newton's minimization method to find the desired Fermi energy.
To validate the approach, we conduct an extensive study calculating the Fermi energy of over 20,000 bulk materials, and we discuss the implications of an incorrect Fermi-energy determination in the electronic properties of these materials.
As an aside, we also show how the standard approaches of Gaussian and Fermi-Dirac smearing can be considered as equivalent provided a renormalization in the smearing width by a factor $\sim$2.565 is accounted for.

\section{Smearing formalism}

\subsection{Broadening and entropy}

Focusing on a general formulation to introduce and gauge different schemes used to smooth the Fermi discontinuity in metals, De Vita introduced the concept of generalized free energy~\cite{de_vita_energetics_1992}.
This is accomplished by introducing an additional term $S[\{f_i\}]$ in the energy functional to make it variational with respect to the occupation numbers:
\begin{equation}
\begin{split}
    A[\sigma;& \{\psi_i\}, \{f_i\}] =  \sum_i f_i \bra{\psi_i} \hat H \ket{\psi_i} \\ 
    &+ \int \left( \mathcal{E}_\text{xc}(\VEC r) - \mathcal{V}_\text{xc}(\VEC r) - \frac{1}{2} \mathcal{V}_\text{H}(\VEC r) \right) \rho(\VEC r) \text d \VEC r \\
    &- \sigma S[\{f_i\}] +  \mu \Big(N-\sum_i f_i\Big) + \sum_i \lambda_i \big(1-\braket{\psi_i | \psi_i} \big) ,
\end{split}
\end{equation}
where $n(\VEC r) = \sum_i f_i \psi_i^*(\VEC r) \psi_i(\VEC r)$, $f_i$ are the occupation functions, $\mathcal E_\text{xc}$ and $\mathcal{V}_\text{xc}$ are the exchange-correlation energy density and potential, respectively, and $\mathcal V_\text{H}$ is the Hartree potential.
The generalized ``entropy'' $ S[\{f_i\}]$, which is a function of the occupation only, is added to satisfy the variational requirement; $\sigma$ is analogue to a temperature, and $\mu$ and $\lambda_i$ are Lagrange multipliers used to impose charge conservation and normalization of the orbitals.

Imposing the stationary requirements of having a self-consistent minimum on this generalized functional, the following condition must be satisfied:
\begin{equation}\label{eq:equilibrium_condition}
    \frac{\partial A}{\partial f_i} = 0 \quad \Rightarrow  \quad \frac{\partial S}{\partial f_i} = \frac{\epsilon_i - \mu}{\sigma} ,
\end{equation}
which provides the fundamental link between the generalized entropy $S$, occupations $f_i$, and the expectation values $\epsilon_i$ of the Hamiltonian.
Furthermore, instead of following the statistical mechanics approach of maximizing the physical entropy to determine the equilibrium occupation function, in this approach the occupation function can be chosen arbitrarily, and for each possible choice, an entropy is then derived from the minimization requirements (i.e., from Eq.~\eqref{eq:equilibrium_condition}), as discussed below.

The arbitrary occupation function $f$ is chosen to be written as an integral of a broadening function
\begin{equation}\label{eq:occupation_function}
f(x) = \int_{-\infty}^x \tilde \delta (\epsilon)\text  d\epsilon ,
\end{equation}
where the broadening function $\tilde \delta(\epsilon)$ is normalized to 1 and $x = \frac{\mu-\epsilon}{\sigma}$. 
This relation provides an operative definition of the occupations deriving from the fictitious temperature as an integrated broadening, with the full freedom to choose the broadening function, as long as the usual physical constraints on the occupancies are satisfied.
In a non-interacting description, it is natural to choose $S$ as a linear combination of single-particle terms:
\begin{equation}
S[\{f_i\}] = \sum_i S_i = \sum_i S(f_i) , 
\end{equation}
where $S(f)$ is a function that is determined by manipulating and integrating Eq.~\eqref{eq:equilibrium_condition};
\begin{equation}\label{eq:entropy_new}
\frac{\text d S}{\text d f} =
- x \,\, \Rightarrow \,\,
\frac{\text d S}{\text d x} = - x \frac{\text d f}{\text d x} \,\, \Rightarrow \,\,
S(x) = -\int_{-\infty}^{x} \epsilon \tilde \delta(\epsilon) \text d \epsilon .
\end{equation}
The above equation then provides a connection between the choice of broadening function and the actual form of the entropy; see also Ref.~\onlinecite{marzari_ab-initio_1996} for an in-depth discussion.
We will discuss the various choices for the broadening function $\tilde \delta$ further ahead.

\subsection{Smeared density of states}

Smearing schemes were first employed to improve the self-consistent convergence in the presence of level crossing and the accuracy of Brillouin-zone sampling by Fu and Ho~\cite{fu_first-principles_1983} and Needs, Martin and Nielsen~\cite{needs_total-energy_1986}.
They were based on the idea of broadening the exact density of state to partially include the contribution of neighboring regions surrounding each $k$-point of a finite set.
The density of states at zero temperature (unsmeared density of states) is given by
\begin{equation}
    n(\epsilon) = \sum_{i \VEC k} \delta(\epsilon - \epsilon_{i \VEC k})  ,
\end{equation}
where $\epsilon_{i\VEC k}$ are the eigenvalues of the Kohn-Sham system with $i$ being the band index and $\VEC k$ the wave vector sampling the first Brillouin Zone.
One then defines a smeared density of states $\tilde n $ through a convolution of the zero temperature density with a broadening function $\tilde \delta$, which is usually a smoother approximation to Dirac's delta:
\begin{equation}\label{eq:smeared_density}
    \tilde n(\epsilon) = \int^{\infty}_{-\infty}  \frac{1}{\sigma} \tilde \delta \left( \frac{\epsilon - \epsilon'}{\sigma} \right) n(\epsilon')  \text{d}\epsilon' ,
\end{equation}
where $\sigma$ corresponds to the smearing parameter.
As $\sigma \rightarrow 0 $, the unsmeared density of states is recovered.

The connection between a smeared density of states and an electronic temperature has been made explicitly by De Gironcoli~\cite{de_gironcoli_lattice_1995}.
The new smeared density of states defines a smeared total energy through the band index sum $\sum_i \epsilon_i$ that can be written as
\begin{equation}\label{eq:smeared_total_energy}
    \tilde E = \int^\mu_{-\infty} \epsilon \tilde n (\epsilon)  \text d \epsilon ,
\end{equation}
where $\mu$ is the chemical potential.
By replacing Eq.~\eqref{eq:smeared_density} in the equation above and inverting the order of the integration, one obtains the following~\cite{de_gironcoli_lattice_1995}
\begin{equation}\label{eq:total_energy_entropy_term}
\begin{split}
    \tilde E = &
    \int_{-\infty}^\infty n(\epsilon') \text d \epsilon'
    \int_{-\infty}^\mu \epsilon  \frac{1}{\sigma} \tilde \delta \left( \frac{\epsilon - \epsilon'}{\sigma}\right) \text d \epsilon
 \\
 = &
    \int_{-\infty}^\infty n(\epsilon') \text d \epsilon'
    \int_{-\infty}^\mu \sigma \left( \frac{\epsilon - \epsilon' + \epsilon'}{\sigma} \right)  \frac{1}{\sigma} \tilde \delta \left( \frac{\epsilon - \epsilon'}{\sigma}\right) \text d \epsilon 
 \\
  = &
    \int_{-\infty}^\infty \epsilon' n(\epsilon') \text d \epsilon'
    \int_{-\infty}^\mu \frac{1}{\sigma} \tilde \delta \left( \frac{\epsilon - \epsilon'}{\sigma}\right) \text d \epsilon \\
    &+ 
    \sigma
    \int_{-\infty}^\infty n(\epsilon') \text d \epsilon'
    \int_{-\infty}^\mu  \left( \frac{\epsilon - \epsilon'}{\sigma} \right)  \frac{1}{\sigma} \tilde \delta \left( \frac{\epsilon - \epsilon'}{\sigma}\right) \text d \epsilon 
 \\
 = &    
    \int^{\infty}_{-\infty} \epsilon' \tilde n (\epsilon')  \text d \epsilon'
    \int^{\frac{\mu - \epsilon'}{\sigma} }_{-\infty} \tilde \delta (x)  \text dx \\
    &+ \sigma \int^{\infty}_{-\infty}  \tilde n (\epsilon')  \text d \epsilon'
    \int^{\frac{\mu - \epsilon'}{\sigma}}_{-\infty} x \tilde \delta (x)  \text d x , \\
\end{split}
\end{equation}where in the second line $\frac{\epsilon'}{\sigma}$ was added and subtracted, followed by the variable transformation $x = \frac{\epsilon-\epsilon'}{\sigma}$.
By comparing integrals in $x$ on the last line of the above equation with Eqs.~\eqref{eq:occupation_function} and \eqref{eq:entropy_new}, one obtains:
\begin{equation}\label{eq:total_energy_link}
\begin{split}
\tilde E = &    
    \int^{\infty}_{-\infty} \epsilon' n (\epsilon')  
    f\left( \frac{\mu - \epsilon'}{\sigma} \right) \text d \epsilon' \\
    &+ \sigma \int^{\infty}_{-\infty}   n (\epsilon')  S\left( \frac{\mu-\epsilon'}{\sigma}\right) \text d \epsilon'
    \\
    = &
 \sum_{i\VEC{k} } \epsilon_{i \VEC k} f\left( \frac{\mu - \epsilon_{i \VEC k}}{\sigma} \right)
    - \sigma \sum_{i \VEC k} S\left( \frac{\mu - \epsilon_{i \VEC k}}{\sigma}  \right) . 
\end{split}
\end{equation}
Equation~\eqref{eq:total_energy_link} shows that the broadening of the density of states is equivalent to the addition of an entropic term to the total energy~\cite{de_gironcoli_lattice_1995}.

In particular, one can exploit Eqs.~\eqref{eq:total_energy_link} and~\eqref{eq:entropy_new} to remove the error introduced by the smearing to recover the total energy of the unsmeared system~\cite{de_vita_ab_1991}. 
Following De Gironcoli~\cite{de_gironcoli_lattice_1995}, let us expand the unsmeared density of states in powers of $\epsilon$:
\begin{equation}
 n(\epsilon') 
 = \frac{1}{k!} \sum_{k=0}^\infty n^{(k)}(\epsilon) (\epsilon' - \epsilon)^k  ,
\end{equation}
where $n^{(k)}(\epsilon) = \frac{\text d^k n(\epsilon)}{\text d \epsilon^k} $.
Then, the smearing entropy can be expanded in powers of the smearing temperature $\sigma$ as in
\begin{equation}\label{eq:entropy_expansion}
S= \sum_{k=0}^\infty c_k n^{(k-1)}(\mu) \sigma^k ,
\end{equation}with the introduction of the coefficients 
\begin{equation}\label{eq:entropy_coefficient}
c_k = (-1)^{k+1} \frac{1}{k!} \int^\infty_{-\infty} \epsilon^{k+1} \tilde \delta (\epsilon) \text d \epsilon 
\end{equation}
and the generalization $n^{(-1)}(\mu) = N$.
This result shows that the entropy has no zero-order term in $\sigma$ for even broadening functions  $\tilde \delta$.
Therefore, the first term in the entropy series is linear in $\sigma$, which yields a quadratic dependency for the total free energy. Advanced smearing techniques aim at broadening functions that remove this quadratic dependence.

\subsection{Broadening function $\tilde \delta$}

\begin{figure}[tb]
  \setlength{\unitlength}{1cm}
  \newcommand{\boxsize}{0.3}
  
  \begin{picture}(9,10.2)
     \put( 0.1, 5.70){ \includegraphics[width=8.7cm, trim={0.0cm 2.2cm 0cm 0cm},clip=true]{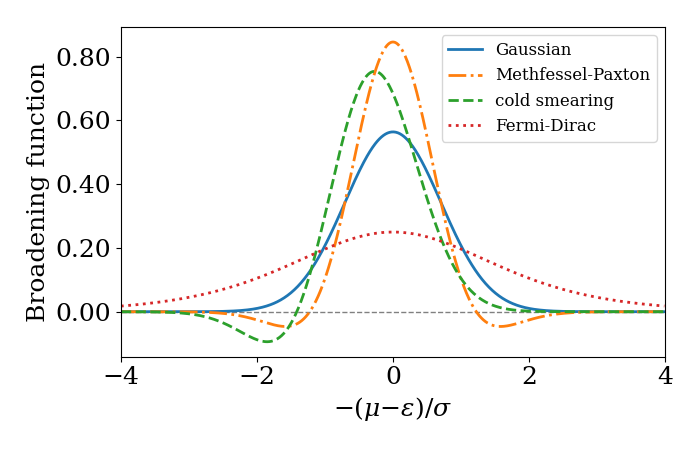} }
     \put( 0.1, 0.00){ \includegraphics[width=8.7cm, trim={0.0cm 0.0cm 0cm 0cm},clip=true]{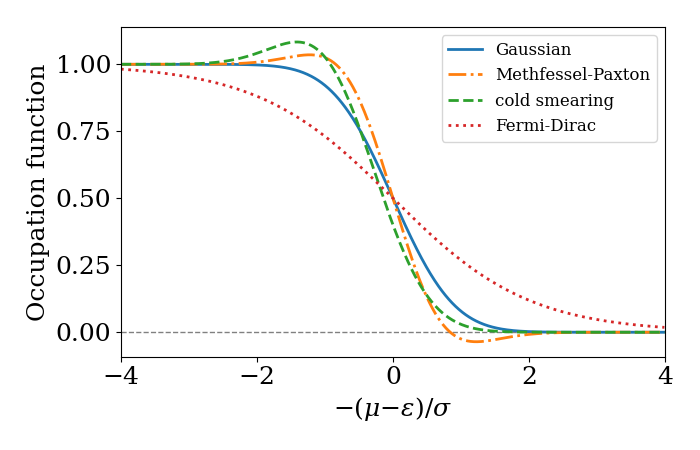} }
     \put(-0.1, 9.6){ \makebox(\boxsize,\boxsize){(a)} }
     \put(-0.1, 4.9){ \makebox(\boxsize,\boxsize){(b)} }
  \end{picture}
  \caption{\label{fig:occupations}
(a) Broadening functions $\tilde\delta(x)$ of the various smearing methods.
For the same smearing, the Fermi-Dirac broadening function (dotted) decays very slowly in comparison to the other approaches.
(b) Resulting occupation function $f(x)$ associated with each of the smearing methods.
The Gaussian (solid) and Fermi-Dirac occupations are monotonic functions.
Meanwhile, the Methfessel-Paxton (dash-dotted) and cold-smearing (dashed) occupation functions are non-monotonic.
In addition, the Methfessel-Paxton occupation function is non-positive definite.
  }
\end{figure}

We now know that the different choices for smearing can be tracked down to the choice of broadening function.
For all smearing methods, the broadening function must satisfy the condition of normalization and we must impose for consistency that the entropy vanishes at zero temperature:
\begin{equation}\label{eq:broadening_conditions}
\begin{split}
&\text{I:} \quad \int^{\infty}_{-\infty} \tilde \delta(\epsilon) \text d \epsilon =1 \quad \rightarrow \quad \text{normalization} \\
&\text{II:} \quad \int^{\infty}_{-\infty} \epsilon \tilde \delta(\epsilon) \text d \epsilon =0 \quad \rightarrow \quad  S(0) = 0 .
\end{split}
\end{equation}
In the following, we discuss the various choices of the broadening function, their features, and qualities.

\emph{Fermi-Dirac smearing:} The broadening function
\begin{equation}
    \tilde \delta (x) = \frac{1}{2\cosh(x) + 2} 
\end{equation}
with $x = \frac{\mu - \epsilon}{\sigma}$ leads, via Eq.~\eqref{eq:entropy_new}, to the well-known entropy
\begin{equation}
S(x)=-f(x) \ln f(x) -\big(1-f(x)\big) \ln\big(1-f(x)\big), 
\end{equation}
where $f(x) = \frac{1}{e^{-x}+1}$ is the Fermi-Dirac distribution function.
We are considering an occupation per spin such that the broadening function integrates to 1 as appropriate for spin-polarized calculations.
The Fermi-Dirac broadening function was considered to have slowly-decaying tails that make energy integrations costly, see Fig.~\ref{fig:occupations}(a).
The most common alternatives are the Gaussian, the Methfessel-Paxton, and the cold smearing broadening functions, which are also represented in Fig.~\ref{fig:occupations}(a).
Figure~\ref{fig:occupations}(b) shows their corresponding occupation functions.

\emph{Gaussian smearing:} This method uses the Gaussian  broadening function
\begin{equation}
\tilde\delta(x) = \frac{1}{\sqrt \pi} e^{- x^2}.
\end{equation}
This function has fast decaying tails, as seen in Fig.~\ref{fig:occupations}(a), which makes the Gaussian smearing a common candidate in practical calculations~\cite{fu_first-principles_1983,needs_total-energy_1986}.
The resulting smearing entropy is
\begin{equation}
S(x) = \frac{1}{2\sqrt \pi}  e^{- x^2} .
\end{equation}

\begin{figure}[tb]
  \setlength{\unitlength}{1cm}
  \newcommand{\boxsize}{0.3}
  
  \begin{picture}(9,10.0)
      \put( 0.1, 5.40){ \includegraphics[width=8.7cm, trim={0.4cm 2.099cm 0cm 0cm},clip=true]{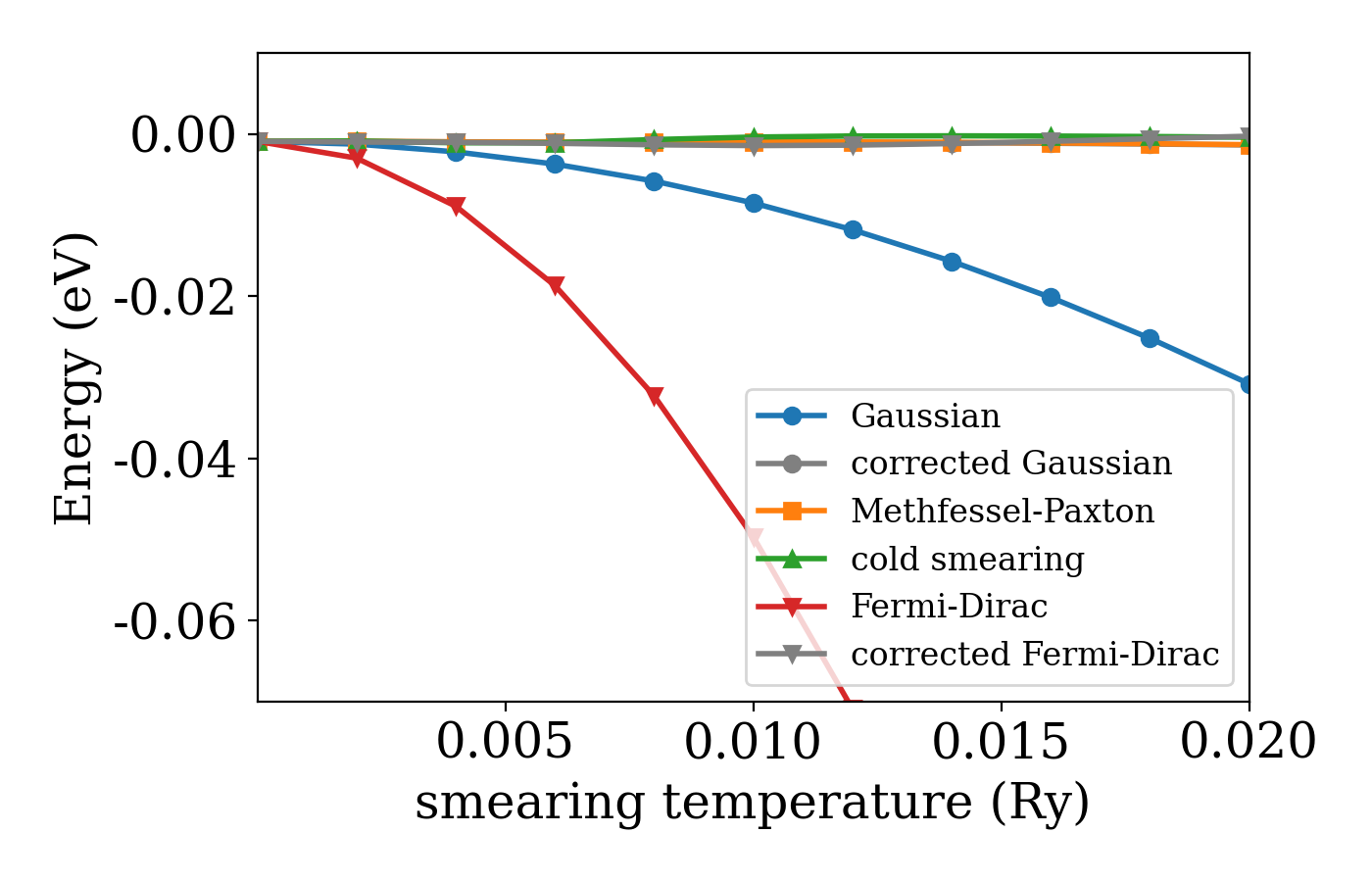} }
     \put( 0.1, 0.00){ \includegraphics[width=8.7cm, trim={0.4cm 0.5cm 0cm 0cm},clip=true]{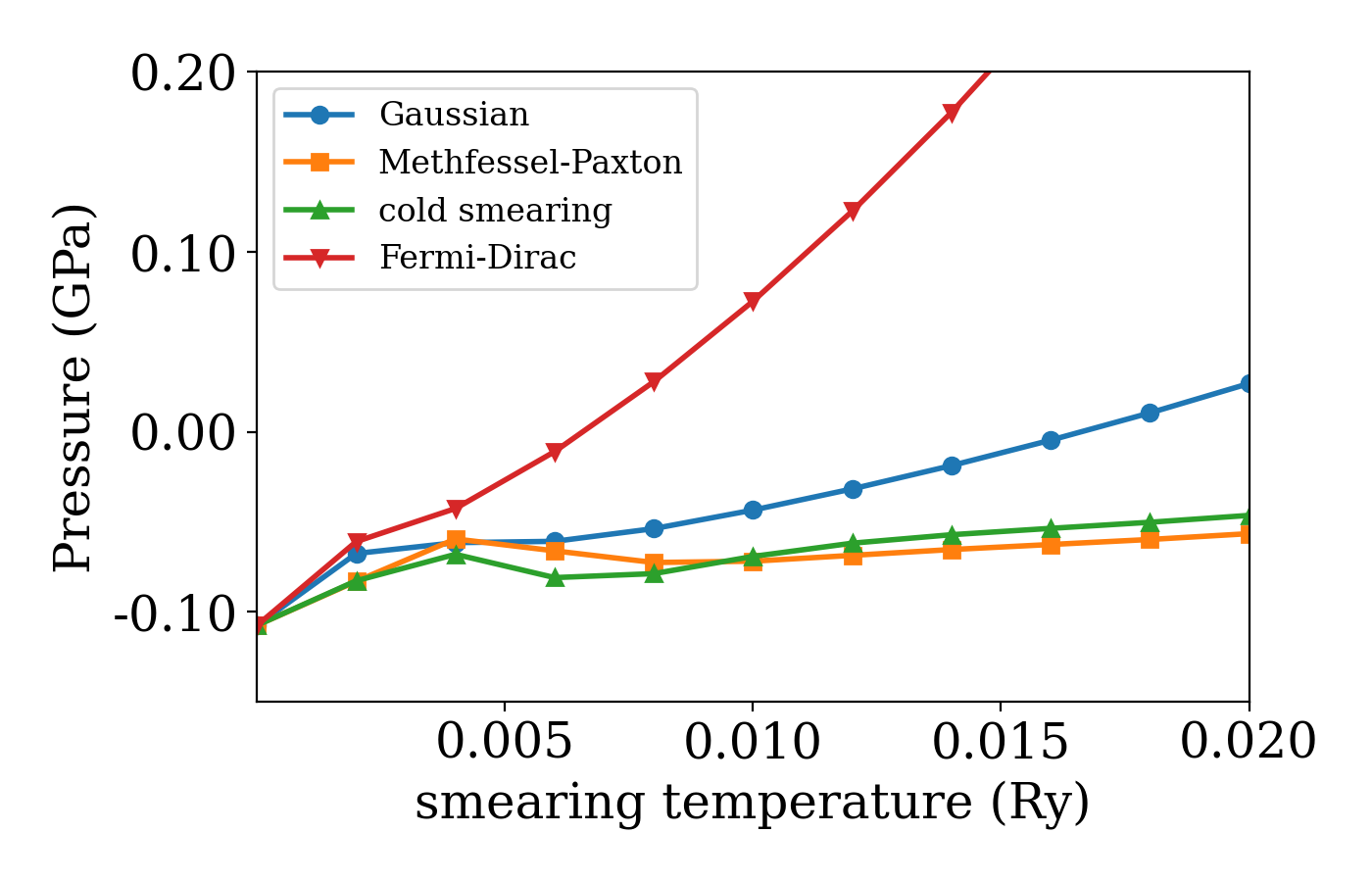} }
     \put(-0.1, 9.4){ \makebox(\boxsize,\boxsize){(a)} }
     \put(-0.1, 4.8){ \makebox(\boxsize,\boxsize){(b)} }
     \Large
  \end{picture}
  \caption{\label{fig:energy_gaussian}
(a)
Free energy as a function of the smearing temperature for bulk Al.
The free energies for the Gaussian (blue circle) and Fermi-Dirac (red down triangle) smearings have a quadratic dependency on the smearing temperature.
The free energies of the Methfessel-Paxton (orange square) and cold (green up triangle) smearings are independent of the smearing temperature up to the third and second power, respectively.
As discussed by De Vita and Gillian~\cite{de_vita_ab_1991}, we can account for the entropic contribution introduced by the Gaussian and Fermi-Dirac smearings to estimate the unsmeared energy with the ``corrected energy'' $\frac{E+A}{2}$ that is shown in gray.
Notice that Methfessel-Paxton, cold, and the corrected Gaussian and Fermi-Dirac agree among themselves in this range of energy.
(b)
Pressure as a function of the smearing temperature.
The Gaussian and Fermi-Dirac smearings affect can strongly affect the pressure in the system due to the expansion of the electron gas. 
A $14\times14\times14$ $k$-point mesh is used; at very small smearing this mesh even might be insufficient to integrate correctly the stress tensor.
  }
\end{figure}

Based on the Taylor expansion of $S$ in terms of $\sigma$ given in Eq.~\eqref{eq:entropy_expansion}, we determine the expansion of the total free energy around $\sigma=0$:
\begin{equation}\label{eq:energy_expansion2}
\begin{split}
A(\sigma) =& A(0) + \sigma (-S)|_{\sigma=0} + \frac{1}{2}\sigma^2 \frac{\text d (-S)}{\text d \sigma}|_{\sigma = 0} + \mathcal O(\sigma^3)\\
=& E(0) - \frac{1}{2}\gamma \sigma^2 + \mathcal O(\sigma^3),
\end{split}
\end{equation}
where we have exploited the fact that at the self-consistent minimum
\begin{equation}\label{eq:entropy_variation}
\frac{\partial A}{\partial \sigma} = \frac{\text d A}{\text d \sigma} = - S(\sigma)
\end{equation}
(since the partial derivatives of the free energy with respect to orbitals and occupancies are zero).
Note that at zero temperature, the total energy $E(0)$ and the free energy $A(0)$ coincide.
Since $A=E-\sigma S$, the dependence of $E$ on $\sigma$ is also determined if $S$, correct up to second order, is introduced in the previous equation:
\begin{equation}\label{eq:total_energy2}
E(\sigma) = E(0) + \frac{1}{2}\gamma \sigma^2 + \mathcal O(\sigma^3).
\end{equation}
From Eqs.~\eqref{eq:energy_expansion2} and \eqref{eq:total_energy2}, one can estimate the zero temperature limit of the total and free energies from the calculations at finite temperatures. 
This entropy-corrected estimate $E_0$~\cite{de_vita_ab_1991} is thus given by
\begin{equation}\label{eq:energy_correction}
E_0(\sigma) = \frac{E(\sigma)+A(\sigma)}{2} = E(0) + \mathcal O(\sigma^3).
\end{equation}
In Fig.~\ref{fig:energy_gaussian}(a), we plot for bulk aluminum the quadratic dependence of the total free energy on $\sigma$ for Fermi-Dirac and Gaussian smearings; the entropy-corrected estimate $E_0$ is also shown.

The entropy-corrected force acting upon an ion at $\VEC R$ is given by~\cite{de_vita_energetics_1992}:
\begin{equation}\label{eq:atomic_forces}
    \VEC{f}_0 = -\frac{\text d E_0}{\text d \VEC R} = -\frac{\text d A}{\text d   \VEC R} - \frac{1}{2}\sigma \frac{\text d S}{\text d \VEC R} =\VEC f_{HF} -\frac{1}{2}\sigma \frac{\text d \VEC{f}_{HF}}{\text d \sigma},
\end{equation}
where $\VEC f = -\frac{\text d A}{\text d   \VEC R}$ is the Hellmann-Feynman force and we used Eq.~\eqref{eq:entropy_variation} to evaluate $\frac{\text d S}{\text d \VEC R}$ exchanging the order of derivatives with respect to $\sigma$ and $\VEC R$.
The above equation demonstrates that the entropy correction on the forces is no longer a straightforward post-hoc contribution.
Instead, it requires the evaluation of the derivative of the Hellmann-Feynman forces with respect to the ionic temperature.
Therefore, although the correction given by Eq.~\eqref{eq:energy_correction} allows us to use Gaussian or Fermi-Dirac smearings and still recover the total energy of the unsmeared system, one is unable to easily provide Hellmann-Feynman forces consistent with the free energy, which hinders applications to molecular dynamics.

Another issue that afflicts Gaussian and Fermi-Dirac smearings is the expansion of the electron gas that can rapidly increase the system's pressure affecting the material lattice constant, as shown in Fig.~\ref{fig:energy_gaussian}(b).

\emph{Methfessel-Paxton smearing:} 
To try and address some of these problems, Methfessel and Paxton~\cite{methfessel_high-precision_1989} proposed as a different strategy a broadening function that, in the language of free energies discussed above, removes the coupling of the free energy with the smearing temperature.
With this goal, they chose as broadening functions $\tilde\delta(\epsilon)$ the first $N$ terms of Dirac's delta expansion in Hermite polynomials, that is
\begin{equation}
\tilde\delta_N(x) = \sum_{n=0}^{N} A_n H_{2n} e^{-x^2}, 
\end{equation}
where
\begin{equation}
\begin{split}
A_n = & \frac{(-1)^n}{n! 4^n \sqrt \pi}  \quad \text{and} \\
H_{n+1}(x) = & 2 x H_n(x) - 2 n H_{n-1}(x),
\end{split}
\end{equation}
with $H_0 = 1$ and $H_ 1(x) = 2x$. 
By construction, the coefficients $c_k$ (Eq.~\eqref{eq:entropy_coefficient}) of the entropy expansion up to the order $2N + 1$ are zero.
In practice, $N=1$ is used because it eliminates the quadratic (main contribution) and the cubic terms in the free energy (linear and second order in the entropy).
Also, adding more terms would make the $k$-point convergence worse because in the limit $N \rightarrow \infty$, the Methfessel-Paxton broadening function goes back to Dirac's delta.
The independence of the smeared total free energy of the Methfessel-Paxton smearing (with $N=1$) can be seen in Fig.~\ref{fig:energy_gaussian}(b).

The advantage of this approach is that ionic forces and other derivatives of the free energy (such as stress) are consistently calculated from the Hellmann-Feynman theorem without the need for a-posteriori corrections.
This smearing, however, introduces two problems: the occupation function is neither monotonic nor positive definite, as shown in Fig.~\ref{fig:occupations}(b).
In particular, the latter could bring consequences from theoretical or practical points of view: 
the electron charge density is no longer guaranteed to be positive definite, and this would be particularly relevant, e.g., for the LUMO of a molecule becoming occupied upon chemisorption.

\emph{Cold smearing:}
To correct the problem of the negative occupation functions, Marzari \textit{et al.}~\cite{de_vita_energetics_1992,marzari_ab-initio_1996,marzari_thermal_1999} proposed a new broadening function to give rise to what is known as cold smearing, in reference to the low coupling of the free energy with the smearing temperature.
This was achieved by imposing further constraints to the broadening function besides the ones in Eq.~\eqref{eq:broadening_conditions}:
\begin{equation}
\begin{split}
&\text{III:} \quad \int^{\infty}_{-\infty} \epsilon^2\tilde \delta(\epsilon) \text d \epsilon =0 \quad \rightarrow \quad \text{temperature decoupling} \\
&\text{IV:} \quad \int^{x}_{-\infty}  \tilde \delta(\epsilon) \text d \epsilon \geq 0 \quad \rightarrow \quad \text{positive occupations} .
\end{split}
\end{equation}
Condition III makes the first-order coefficient of the entropy expansion in Eq.~\eqref{eq:entropy_expansion} vanish; this means that the leading term of the free-energy dependence on temperature (the quadratic term) is zero.
Condition IV is required to ensure the occupation function of Eq.~\eqref{eq:occupation_function} remains positive.

A broadening function satisfying these conditions is given in Ref.~\onlinecite{marzari_thermal_1999}:
\begin{equation}
\tilde\delta(x) = \frac{1}{\sqrt \pi} (2 - \sqrt{2} x) e^{-\left( x - \frac{1}{\sqrt 2} \right)^2},
\end{equation}
which is also shown in Fig.~\ref{fig:occupations}(a).
The resulting occupation function is presented in Fig.~\ref{fig:occupations}(b).
Similarly to Methfessel-Paxton, the cold smearing occupation function is also non-monotonic.
However, contrary to the former, it is by construction always positive definite.

\subsection{Gaussian vs Fermi-Dirac}

\begin{figure}[tb]
  \setlength{\unitlength}{1cm}
  \newcommand{\boxsize}{0.3}
  \begin{picture}(8.5,15.2)
    \put(0.0, 10.10){
      \put( 0.0, 0.0){ \includegraphics[width=8.7cm, trim={0.0cm 0.0cm 0.0cm 0cm},clip=true]{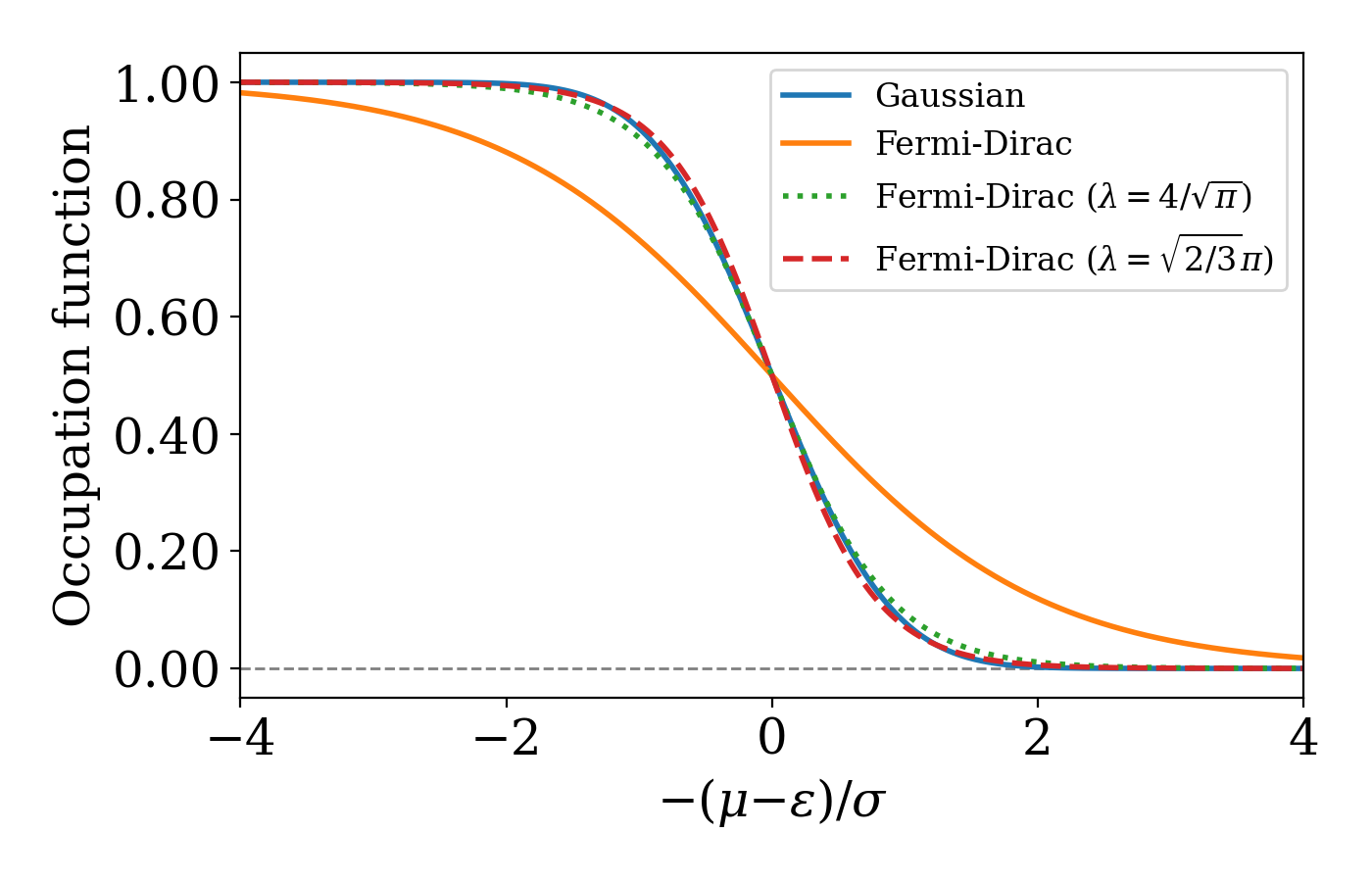} }
      \put( 1.4,1.0){ \includegraphics[width=3.7cm, trim={0.0cm 0.0cm 0.0cm 0cm},clip=true]{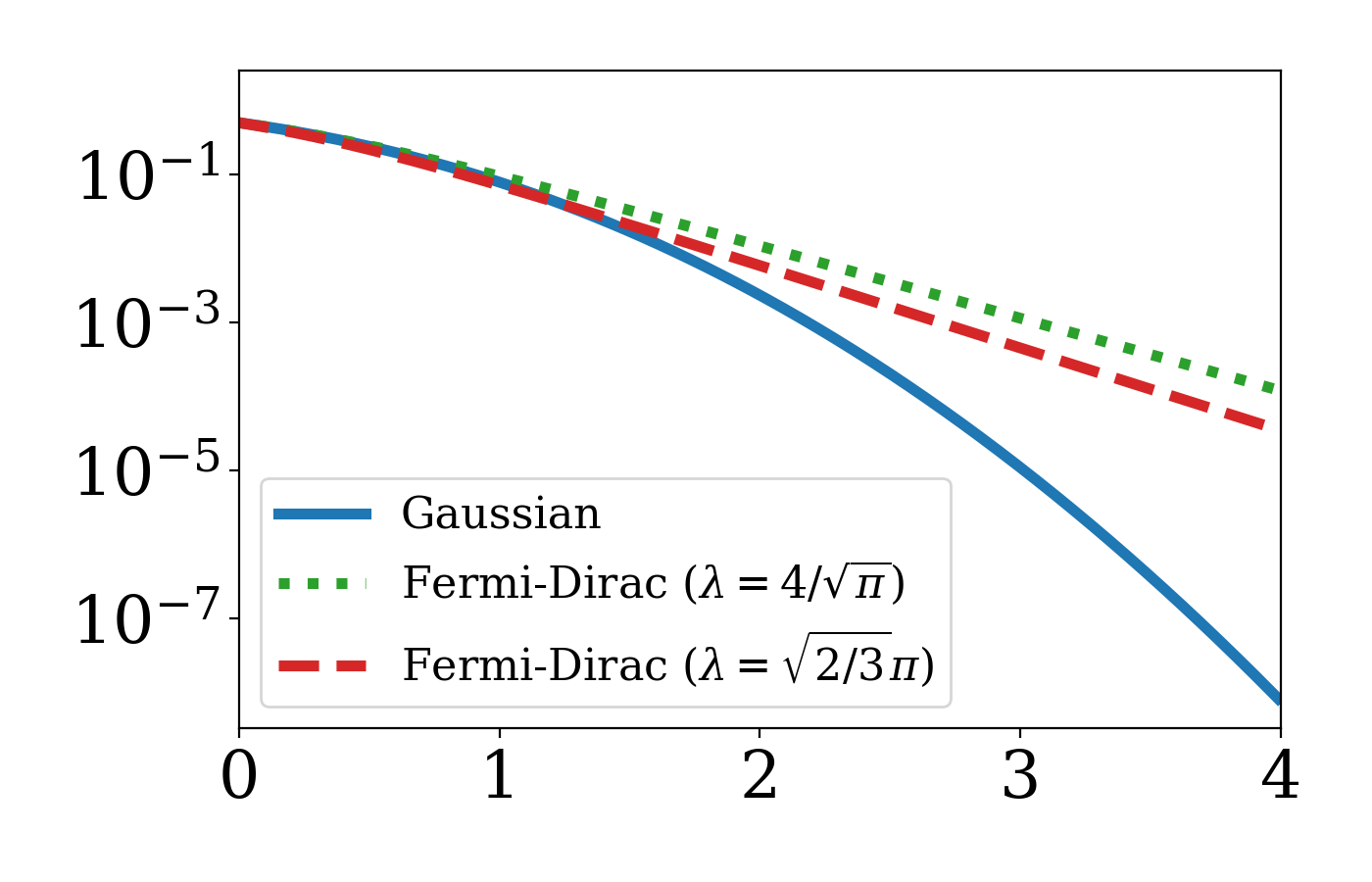} }
    }
    \put( 0.0, 4.85){ \includegraphics[width=8.7cm, trim={0.0cm 0.0cm 0.0cm 0cm},clip=true]{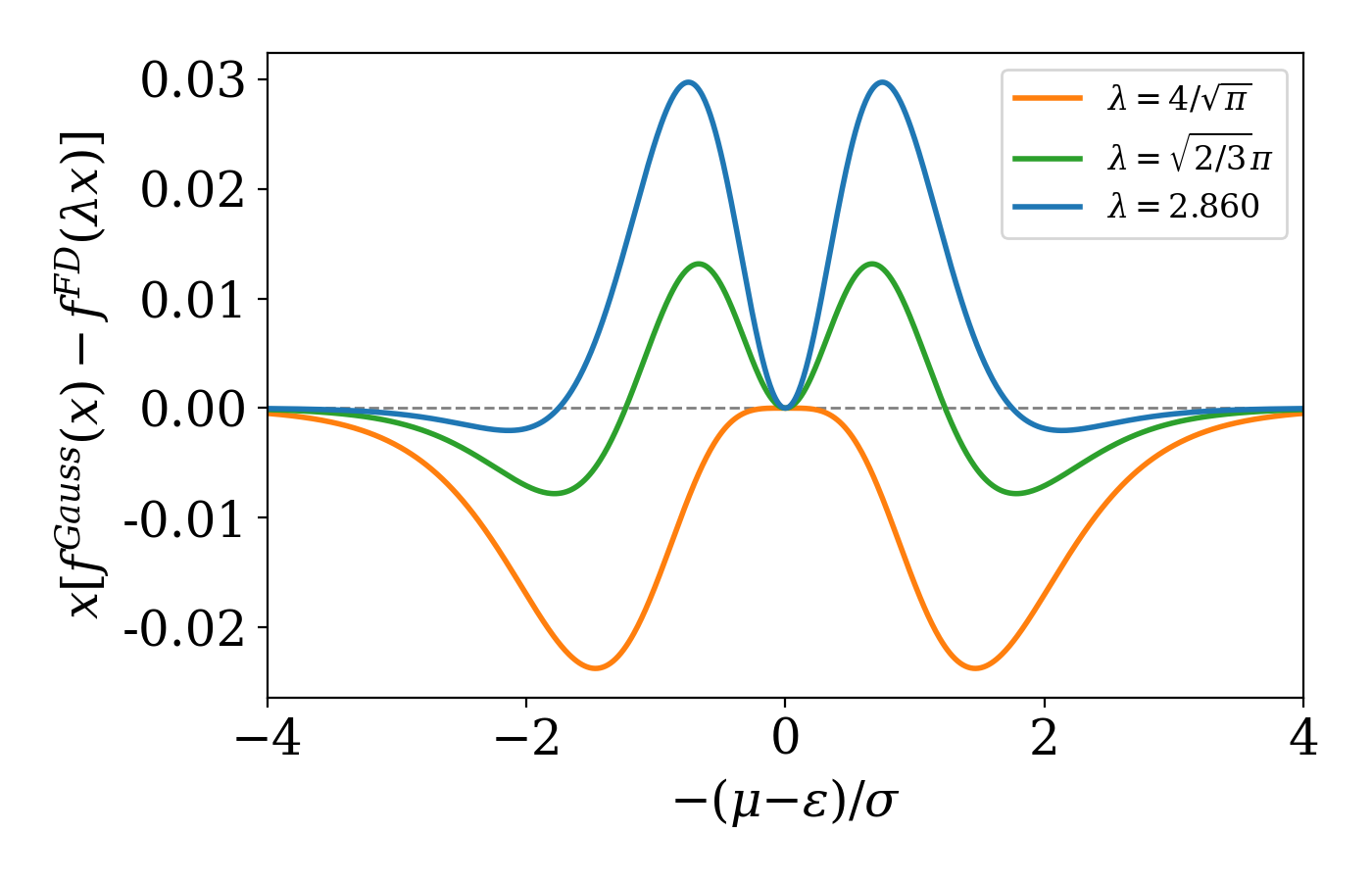} }
    \put( 0.0, -0.1){ \includegraphics[width=8.5cm, trim={0.5cm 0cm 0.5cm 0cm},clip=true]{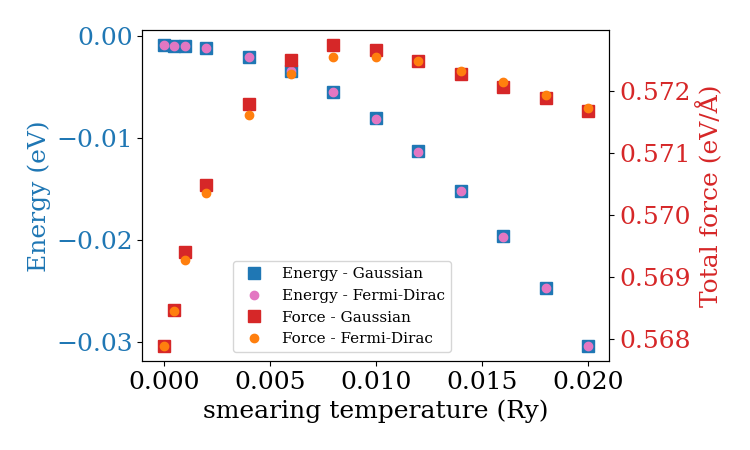} }
    \put(-0.1, 15.0){ \makebox(\boxsize,\boxsize){(a)} }
    \put(-0.1, 9.9){ \makebox(\boxsize,\boxsize){(b)} }
    \put(-0.1, 4.7){ \makebox(\boxsize,\boxsize){(c)} }
  \end{picture}
  \caption{\label{fig:force_vs_degauss_FD_GAUSS}
  Comparison between Gaussian and Fermi-Dirac smearings.
  (a) Fermi-Dirac occupation function $f(x)$ with scaled smearing temperature ($\sigma/\lambda$).
  The inset presents the vertical axis in a log scale demonstrating that the Gaussian occupation decays to zero more rapidly than the Fermi-Dirac one.
  (b) Integrand of Eq.~\ref{eq:occupation_diff}.
  For $\lambda = \sqrt{2/3} \pi$, it integrates to zero in either the negative or positive halves of the horizontal axis. 
  (c) Total free energy and total force as a function of the smearing temperature for bulk Al (with a displaced atom in the unit cell).
  The dots and squares are calculations with Gaussian and Fermi-Dirac smearings, respectively.
  The smearing temperature of the Fermi-Dirac results is scaled down by $\lambda=2.565$.
  }
\end{figure}
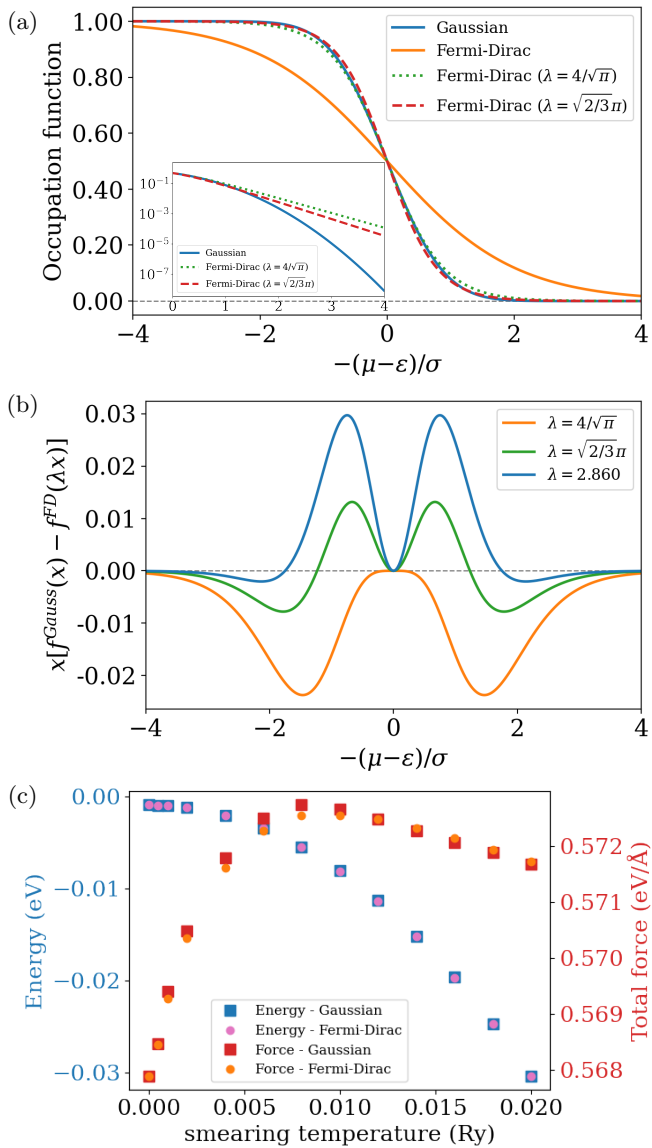

For simplicity, we have so far performed comparisons between Gaussian and Fermi-Dirac smearings at the same temperature $\sigma$.
However, it is important to see if one could match the results of these two approaches by scaling the temperature in one of the simulations.
For example, one could try to match the first derivative of the occupation function at $x=0$.
This is achieved by dividing $\sigma$ in the Fermi-Dirac calculation by $\lambda = 4/\sqrt \pi \sim 2.26$; in Fig.~\ref{fig:force_vs_degauss_FD_GAUSS}(a), one can see how the Gaussian and Fermi-Dirac ($\lambda =4/\sqrt \pi$) occupation curves become similar around  half occupation.
If one computes the total energy using this scaling factor, similar results are obtained, but not an exact match.
Furthermore, the mismatch increases with the smearing temperature.

We showed before that the total free energy is quadratic in $\sigma$ in first-order approximation [see Eqs.~\ref{eq:total_energy_link}, \ref{eq:entropy_expansion}, and \ref{eq:entropy_coefficient}]; this suggests that one should be able to find a temperature scaling to match the Gaussian and the Fermi-Dirac total free energies.
Let us consider the total-free-energy difference between the two methods; we aim to find a $\lambda$ such as:
\begin{equation}
    \tilde E^{\textnormal{Gauss}}(\sigma) - \tilde E^{\textnormal{FD}}\left(\frac{\sigma}{\lambda}\right)= 0 .
\end{equation}
As shown by Eq.~\ref{eq:total_energy_link}, the left-hand side of the equation above can be split in two terms, the difference between the electronic energies and the difference of the entropic contributions.
Searching for the roots of the two terms independently and assuming that the density of states is constant around $\mu$, the electronic-energy term can be written as:
\begin{equation}\label{eq:occupation_diff}
\Gamma(\lambda) = 
\int^{\infty}_{-\infty} x \left[ f^\textnormal{Gauss}(x) - f^\textnormal{FD}(\lambda x )  \right] \text d x 
=  \frac{1}{4} - \frac{\pi^2}{6 \lambda^2} ,
\end{equation}
where we performed the usual change of variable $x = (\mu - \epsilon)/\sigma$.
The Gaussian occupation function is $f^\textnormal{Gauss}(x) = \frac{1}{2}[1 - \text{erf}(x) ] $ and the Fermi-Dirac's  $f^\textnormal{FD}(x) = \frac{1}{e^x + 1}$.
By solving $\Gamma(\lambda) = 0$, we obtain that $\lambda = \sqrt{2/3} \pi \sim 2.565$.
A similar procedure can be followed to show that the entropic differences between the Gaussian and Fermi-Dirac methods also vanishes for this same factor.

The occupation function scaled by $\lambda = \sqrt{2/3} \pi$ is shown in Fig.~\ref{fig:force_vs_degauss_FD_GAUSS}(a).
This curve crosses the Gaussian occupation curve multiple times such that the integrand in Eq.~\ref{eq:occupation_diff} integrates to zero both in the positive and in the negative halves of the domain as shown in Fig.~\ref{fig:force_vs_degauss_FD_GAUSS}(b).
Finally, we performed DFT calculations in bulk Al with two atoms in the unit cell (one displaced out of the equilibrium position to causes non-vanishing forces) using the $\lambda = \sqrt{2/3} \pi$ scaling.
The results are shown in Fig.~\ref{fig:force_vs_degauss_FD_GAUSS}(c), where we can observe a very close match for the free energy between Gaussian and Fermi-Dirac smearing for all smearing temperatures.
The computed forces also show good agreement, although not as close as for the free energy.
This approach is less effective when the density of states varies strongly around the Fermi energy, or in a small-gap insulator.

\subsection{Reciprocal-space sampling and convergence}

\begin{figure}[tb]
  \setlength{\unitlength}{1cm}
  \newcommand{\boxsize}{0.3}
  \begin{picture}(9,10.5)
      \put( 0.0, 5.60){ \includegraphics[width=8.5cm, trim={0.0cm 0.6cm 0cm 0cm},clip=true]{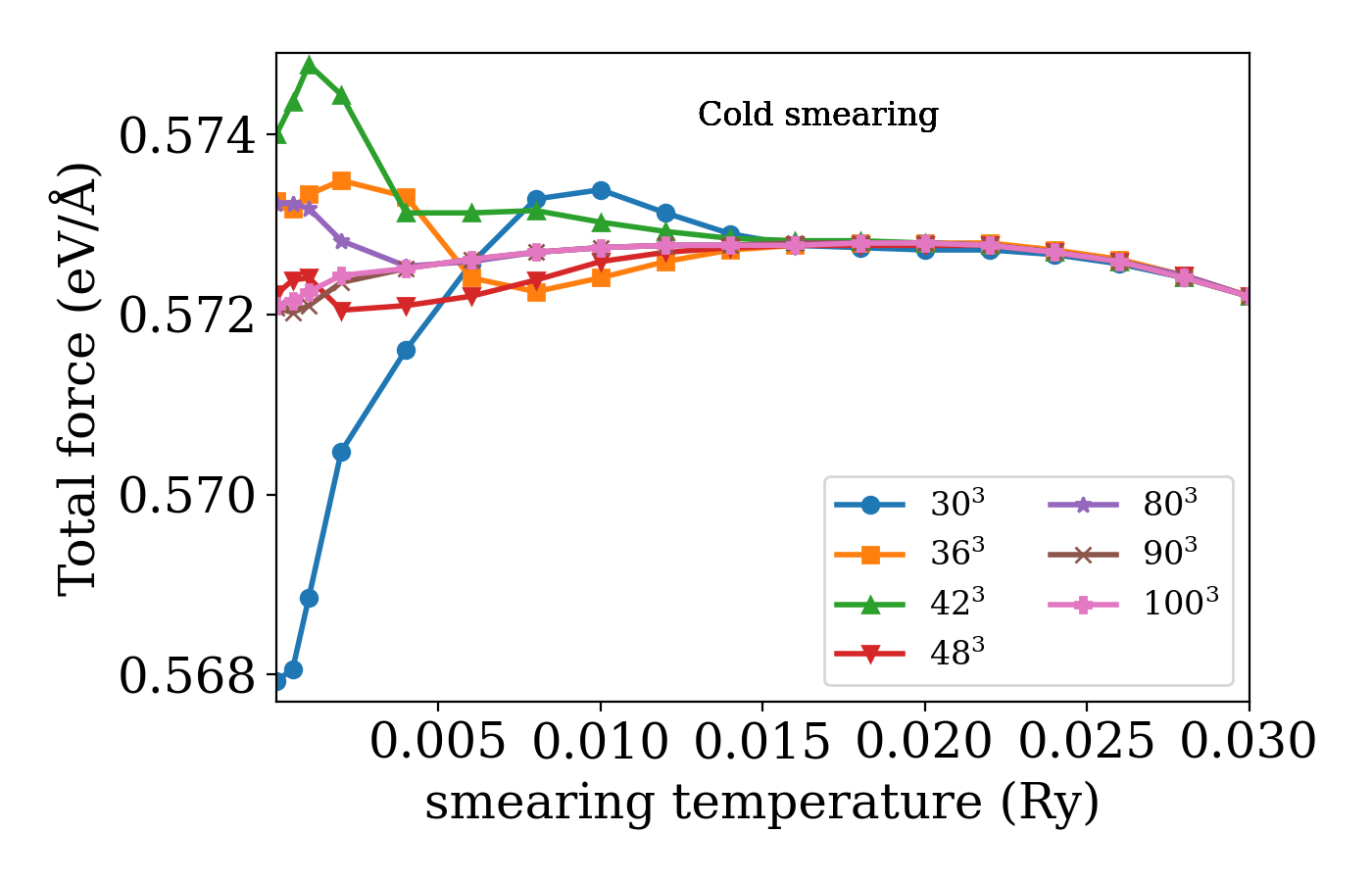} }
     \put( 0.0, 0.00){ \includegraphics[width=8.7cm, trim={0.0cm 0.5cm 0cm 0cm},clip=true]{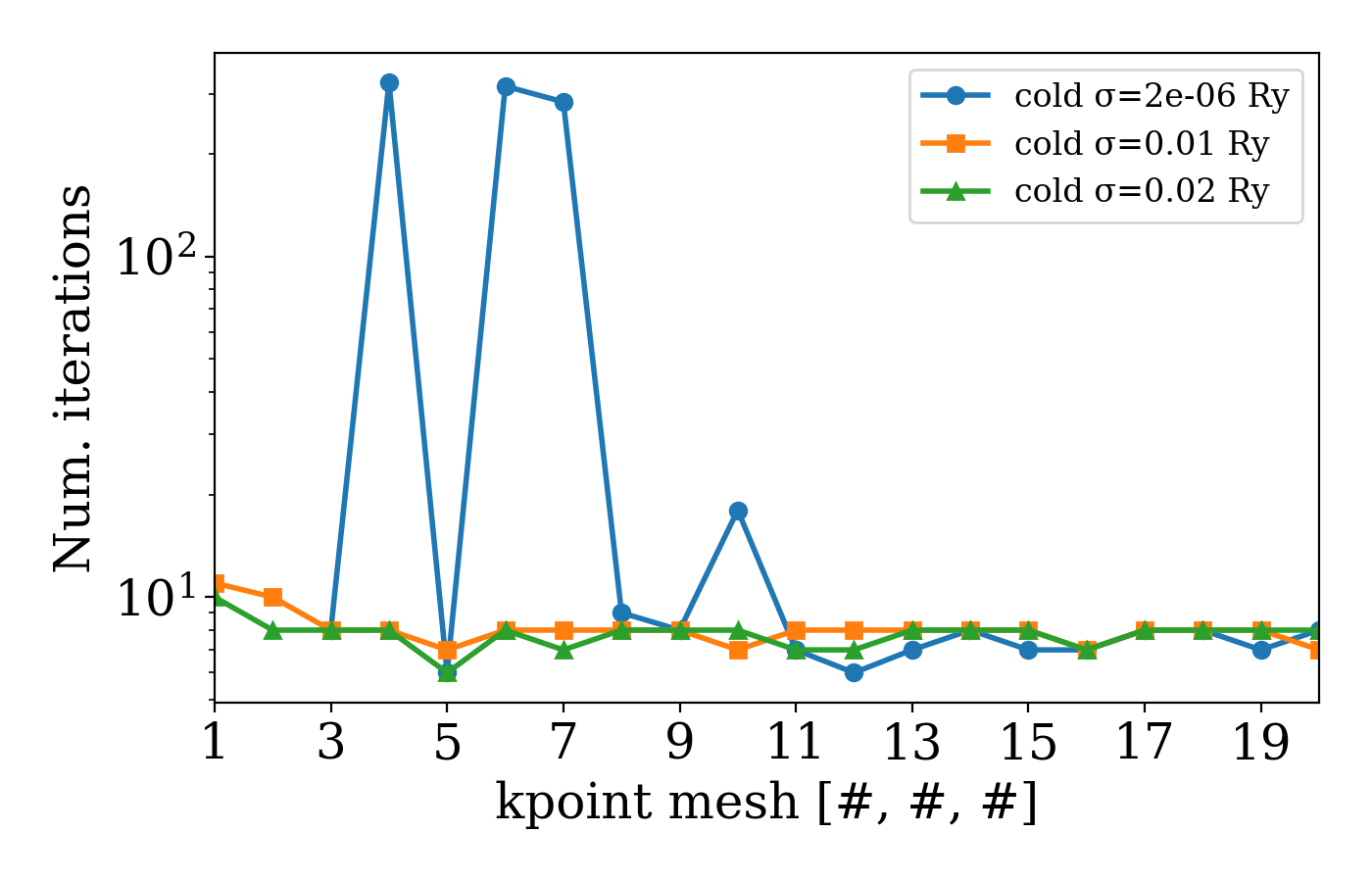} }
     \put(-0.1, 10.2){ \makebox(\boxsize,\boxsize){(a)} }
     \put(-0.1, 4.8){ \makebox(\boxsize,\boxsize){(b)} }
  \end{picture}
  \caption{\label{fig:convergence}
(a) Total force as a function of the smearing temperature $\sigma$ for various k-point mesh.
At very small smearing, the result is not converged even for very fine k-point meshes.
The results converge only for sufficiently high smearing.
(b) Number of iterations needed to achieve self-consistency in the Koln-Sham equations; at small smearings level-crossing instabilities make self-consistency harder to reach (at variance with direct variational minimization~\cite{marzari_ensemble_1997}).
  }
\end{figure}

An important aspect of using smearing technique is to determine the proper range for the smearing $\sigma$; this has also been the subject of recent studies~\cite{lejaeghere_reproducibility_2016,cances_convergence_2021,jorgensen_effectiveness_2021}.
In Fig.~\ref{fig:convergence}(a), we show the total force (the square root of the sum of all of the force components) in a system with two aluminum atoms in the unit cell.
As before, one of the basis atoms was displaced along the [100] direction by 10\% of the nearest neighbour distance.
Note that for very small smearing $\sigma \rightarrow 0$ the forces are not converged even for very fine k-point meshes.
However, for higher smearings, all curves for various k-point mesh converge to the same value.

One must be aware that while smearing makes it possible to converge the reciprocal space sampling, the converged result also depends on smearing.
Smearings such as cold smearing and Methfessel-Paxton are specially designed to mitigate this problem.
However, for high smearings, the high-order dependency of the total free energy on smearing will become relevant.
For example, in Fig.~\ref{fig:convergence}(a), for even higher smearing ($>0.02$ Ry), the converged curves start to deviate as a function of the smearing parameter; thus, the smearing parameter should not be too small, to avoid sampling errors nor too large, to introduce systematic deviations.
For cold smearing, most applications would optimally use a smearing between 0.01-0.02 Ry.
One can decrease the systematic error due to smearing by reducing the smearing parameter ($\sim 0.005$ Ry) but largely increasing the k-point sampling.

In Fig.~\ref{fig:convergence}(b), we illustrate how smearing also improves the self-consistent determination of the Kohn-Sham equation solution when iterative, rather than variational~\cite{marzari_ensemble_1997}, approaches are used.
Occasionally, the self-consistency can take an order of magnitude more iterations.

\subsection{Non-unique chemical potential}

\begin{figure}[tb]
  \setlength{\unitlength}{1cm}
  \newcommand{\boxsize}{0.3}
  \begin{picture}(9,10.2)
      \put( 0.35, 5.60){ \includegraphics[width=8.cm, trim={0.0cm 1.0cm 0cm 0cm},clip=true]{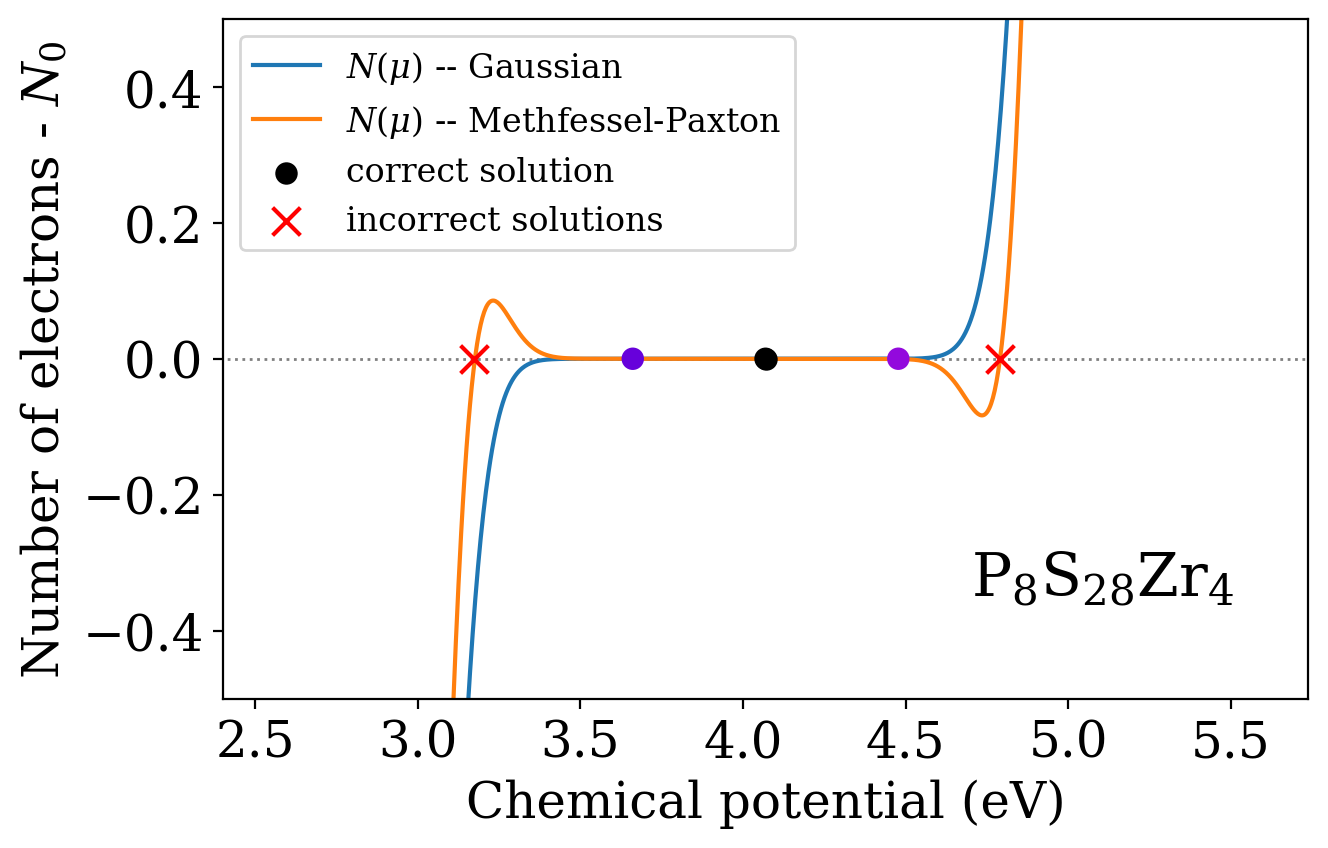} }
     \put( 0.35, 0.00){ \includegraphics[width=8.cm, trim={0.0cm 0.0cm 0cm 0cm},clip=true]{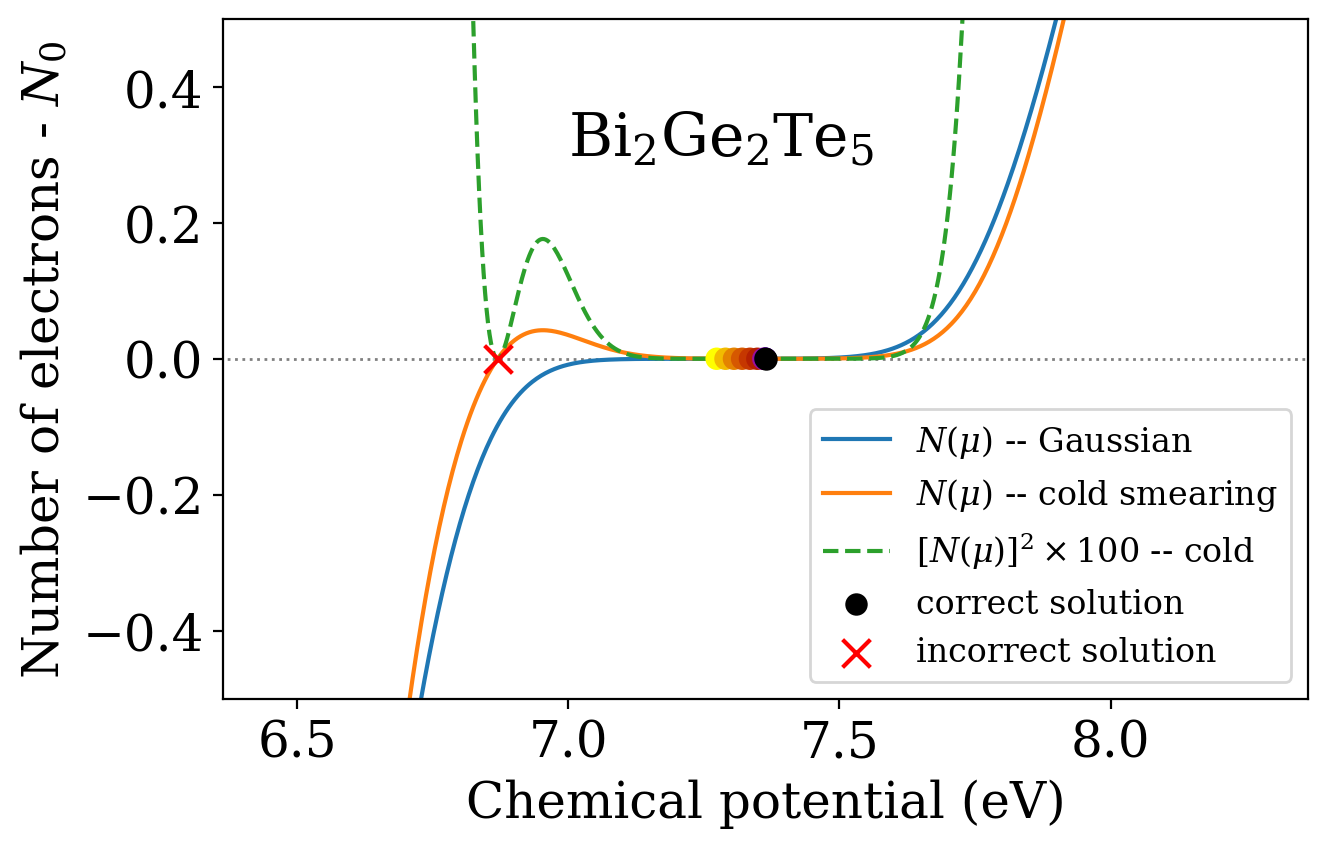} }
     \put(-0.1, 9.85){ \makebox(\boxsize,\boxsize){(a)} }
     \put(-0.1, 4.8){ \makebox(\boxsize,\boxsize){(b)} }
  \end{picture}
  \caption{\label{fig:number_electrons}
Number of electrons as a function of the chemical potential, minus the target number of electrons $N_0$.
(a) Methfessel-Paxton smearing on $\text{P}_8\text{S}_{28}\text{Zr}_4$, which admits up to three nonequivalent Fermi energies.
(b) Cold smearing on $\text{Bi}_2\text{Ge}_2\text{Te}_5$, yielding two solutions.
The blue lines correspond to the Gaussian smearing for reference.
The incorrect solutions are marked by a red ``x" and the correct one by a black dot.
The colored dots are intermediary steps taken by the novel protocol when determining the correct solution.
In (a), the correct solution was found in the prospection step using Gaussian smearing and the bisection method (the purple dots were intermediary bisection steps); in (b), the solution was determined using Newton's minimization on $[N(\mu)]^2$ (green dashed line) starting from the yellow dot obtained in the initial prospection with Gaussian smearing.
  }
\end{figure}

While cold smearing removes negative occupations, it shares with Methfessel-Paxton a non-monotonic occupation function.
As a consequence, the chemical potential can become non-uniquely defined.
The chemical potential is defined by the root of
\begin{equation}\label{eq:number_electrons}
N(\mu) = \sum_{i \VEC k} f \left( \frac{\mu - \epsilon_{i \VEC k}}{\sigma} \right) - N_0 ,
\end{equation}
where $N_0$ is the number of electrons in the system.
In Fig.~\ref{fig:number_electrons}(a) we plot $N(\mu)$ in Eq.~\eqref{eq:number_electrons} for $\text{P}_8\text{S}_{28}\text{Zr}_4$, using Gaussian and Methfessel-Paxton smearings.
We can see that the Methfessel-Paxton curve allows for three non-equivalent solutions, two incorrect solutions at 3.2 and 4.8 eV, and an interval of solutions (analytically, only one at ~4.0 eV) in the plateau between  3.5-4.5 eV.
In Fig.~\ref{fig:number_electrons}(b), the cold smearing for $\text{Bi}_2\text{Ge}_2\text{Te}_5$ results in two possible solutions, the incorrect one at 6.8 eV and the correct one withing the 7.2-7.5 eV plateau.

These multiple chemical potentials usually occur in the case of insulators and semiconductors; despite smearing being meant to be used for metallic and magnetic systems, there are occasions where one may need to use smearing also for semiconductors and insulators, a clear example being the case of high-throughput studies, which aim at calculating the properties of materials without knowing a priori if they are insulators or metals.

\section{Fermi energy determination }

\subsection{Newton's minimization method}

The Fermi energy (or, rather, the chemical potential) is often obtained by the bisection method finding the roots of Eq.~\eqref{eq:number_electrons};
This is a very robust root-finding algorithm that allows determining at least one root of a function if boundaries for the interval containing the root are known.
However, it will not address the possibility of having other roots in the same given interval.
The bisection method for the functions shown in Fig.~\ref{fig:number_electrons}(a) and (b) can return any of the admissible roots, depending only on the interval provided, which typically spans the minimum and maximum of the electronic eigenvalues.
As the chemical potential obtained determines the occupation of the electronic states, an incorrect identification can lead to the incorrect ground state or can be a source of instability in a self-consistent calculation.

\begin{figure}[tb]
  \setlength{\unitlength}{1cm}
  \newcommand{\boxsize}{0.3}
  \begin{picture}(8,8.7)
    \put( 0.0, 0.00){ \includegraphics[width=8.4cm, trim={6cm 0cm 6cm 0cm},clip=true]{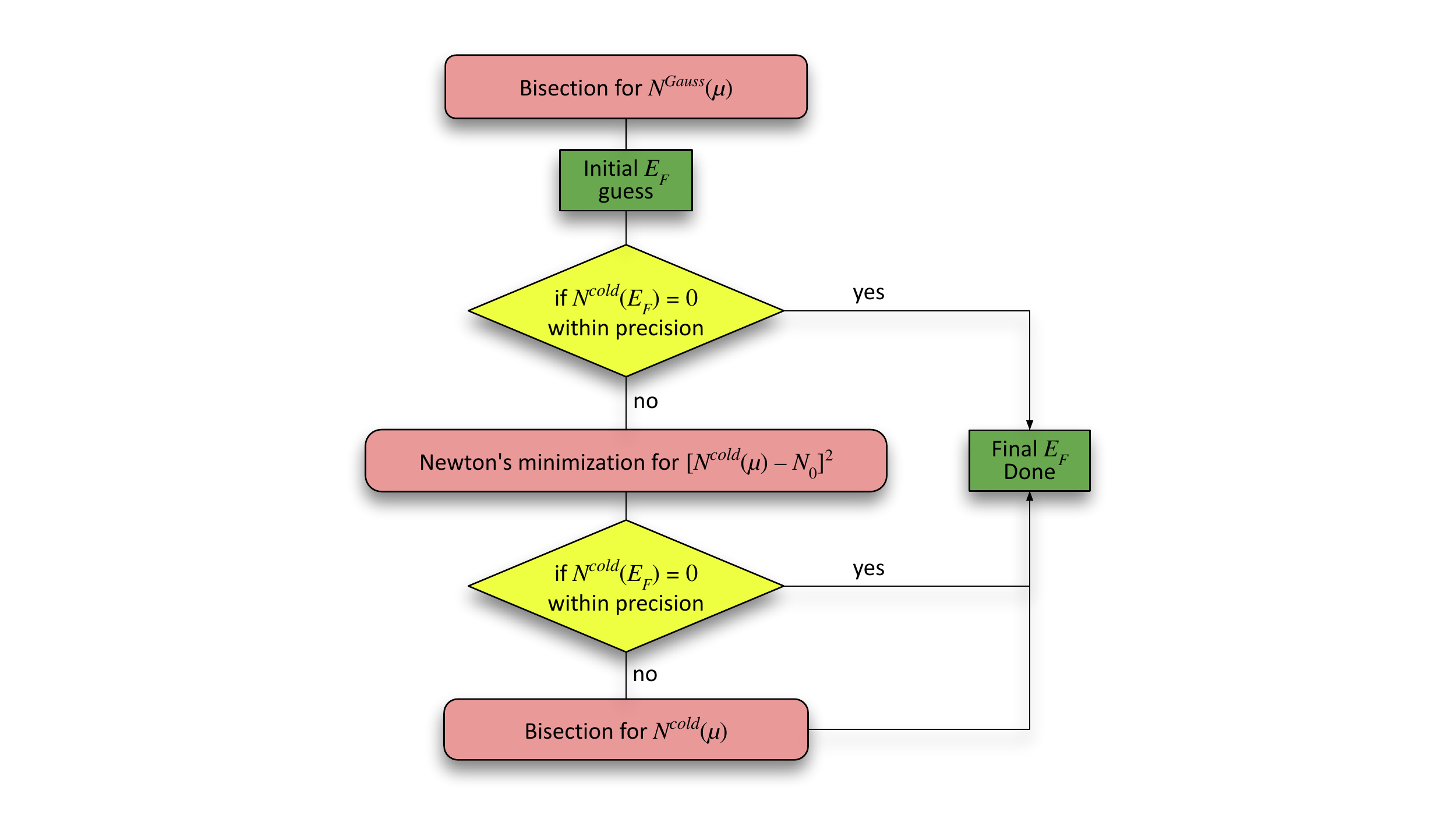} }
  \end{picture}
  \caption{\label{fig:protocol_diagram}
Fermi energy determination algorithm.
First, we compute the chemical potential with Gaussian smearing and the bisection method; this is used as the initial guess for Netwon's minimization algorithm using the desired smearing method (Methfessel-Paxton or cold smearing).
In case the minimization returns a chemical potential that does not yield the correct number of electrons, the bisection method is used for the desired smearing.
  }
\end{figure}

We treat this problem numerically through the following protocol:
we convert the root-finding problem into a minimization one, and start it from the (unique) Fermi energy obtained from a Gaussian broadening function.
With this in mind, we introduce a new function
\begin{equation}\label{eq:new_number_electron}
\tilde N(\mu) =[N(\mu)]^2,
\end{equation}
for which we apply the iterative Newton's minimization algorithm~\cite{press_numerical_2007} to find its minima.
In Newton's minimization, one starts from an initial guess $\mu_0$ and computes consecutive steps determined by
\begin{equation}\label{eq:newtons}
\mu_{t+1} = \mu_t - \frac{\tilde N'(\mu_t)}{\tilde N''(\mu_t)}  .
\end{equation}
In this way, $\mu$ moves towards the closest extremum point of $\tilde N$.
We can force the algorithm to ignore maximum points and always search the minima by considering the absolute value of the denominator in Eq.~\eqref{eq:newtons}.
Note that the function in Eq.~\eqref{eq:new_number_electron} can have multiple minima;
however, this issue can be solved by finding a good guess which is inside the desired valley.
Such a guess is obtained as the chemical potential of a reconnaissance run using Gaussian smearing in combination with the bisection method;
the algorithmic representation of the protocol is given by the diagram in Fig.~\ref{fig:protocol_diagram}.
Newton's minimization algorithm, as given in Eq.~\eqref{eq:newtons}, requires the calculation of the first and second derivatives of the occupation function.
From Eq.~\eqref{eq:occupation_function}, the first derivative is trivially the broadening function of the respective method, and the second derivative can also be straightforwardly derived analytically.

In Fig.~\ref{fig:number_electrons}(a), we show the results of applying such protocol to $\text{P}_8\text{S}_{28}\text{Zr}_4$, using Methfessel-Paxton smearing.
In this example, the protocol finds the correct Fermi energy (black dot) already in the reconnaissance phase.
The purple dots in  Fig.~\ref{fig:number_electrons}(a) are just intermediate steps of the bisection method.
In Fig.~\ref{fig:number_electrons}(b) the protocol is applied for $\text{Bi}_2\text{Ge}_2\text{Te}_5$ using cold smearing.
The initial guess for $E_F$ is represented by the yellow dot.
This time, Newton's minimization is required to further improve the result.
It acts on $[N(\mu)]^2$, which corresponds to the green dashed line (scaled up by 100 in Fig.~\ref{fig:number_electrons}(b) to facilitate the visualization).
The colored dots between the initial step (yellow dot) and the final solution (black dot) are intermediate steps of Newton's minimization algorithm.

\begin{figure}[tb]
  \setlength{\unitlength}{1cm}
  \newcommand{\boxsize}{0.3}
  \begin{picture}(9,5.2)
      \put( 0, 0){ \includegraphics[width=8.cm, trim={0.0cm 0.0cm 0cm 0cm},clip=true]{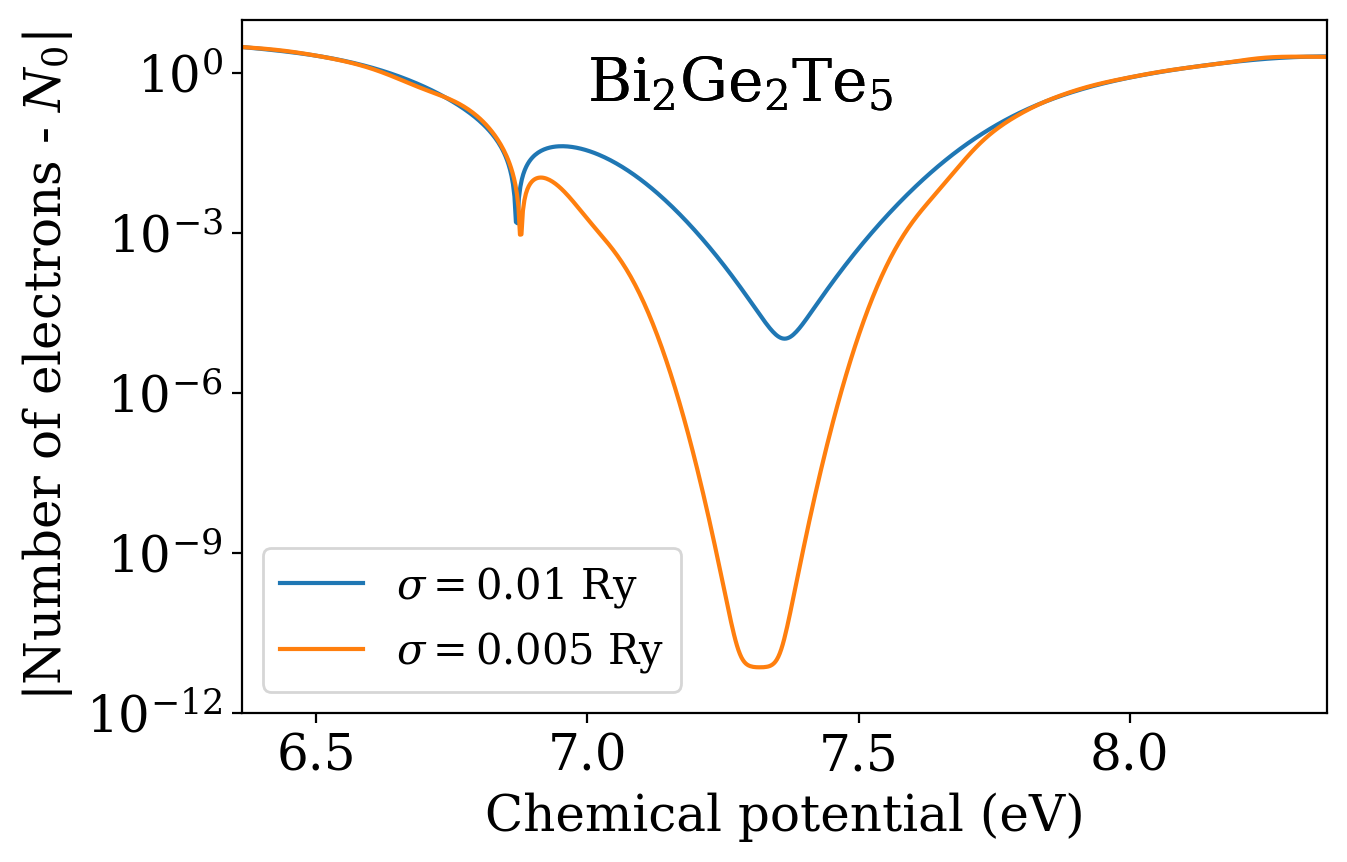} }
  \end{picture}
  \caption{\label{fig:abs_number_electrons}
Absolute value of the number of electrons minus the target number of electrons as a function of the chemical potential.
Calculation for $\text{Bi}_2\text{Ge}_2\text{Te}_5$ using cold smearing for two values of the smearing temperature $\sigma$.
The larger value of $\sigma$ results in an inside-gap minimum (central deep) not very close to zero, thus corresponding to a noninteger number of electrons.
A reduction of $\sigma$ remediates this issue.
}
\end{figure}

When the bandgap is not very large with respect to the smearing temperature $\sigma$, it can occur that $N(\mu)$ does not go exactly to zero within the band gap when using cold smearing.
The absolute value of $N(\mu)$ for $\text{Bi}_2\text{Ge}_2\text{Te}_5$ is displayed in Fig.~\ref{fig:abs_number_electrons} in a log scale for two values of temperature $\sigma$ using cold smearing.
The sharp dips on the left correspond to the undesired chemical potentials, which are true zeros of $N(\mu)$.
The shallow dips at the center represent the solutions inside the band gap.
For $\sigma = 0.01$ Ry, the minimum of the curve is $\sim 10^{-05}$, which implies that the resulting number of electrons is noninteger.
This artifact derives from the fact that the cold-smearing occupation function $f(x)$ is positive definite and tends to 1 a bit more slowly than other methods for large values of $x$.
In these cases, a reduction of the smearing temperature $\sigma$ is recommended.
In our example, halving the smearing temperature makes the minimum of $N(\mu)$ smaller than $10^{-11}$, which corresponds to an integer number of electrons within the numerical precision of typical calculations.

 \subsection{High-throughput validation}

\begin{figure}[tb]
  \setlength{\unitlength}{1cm}
  \newcommand{\boxsize}{0.3}
  \begin{picture}(9,10.2)
      \put( 0.2, 5.40){ \includegraphics[width=8.7cm, trim={0.0cm 1.4cm 0cm 0cm},clip=true]{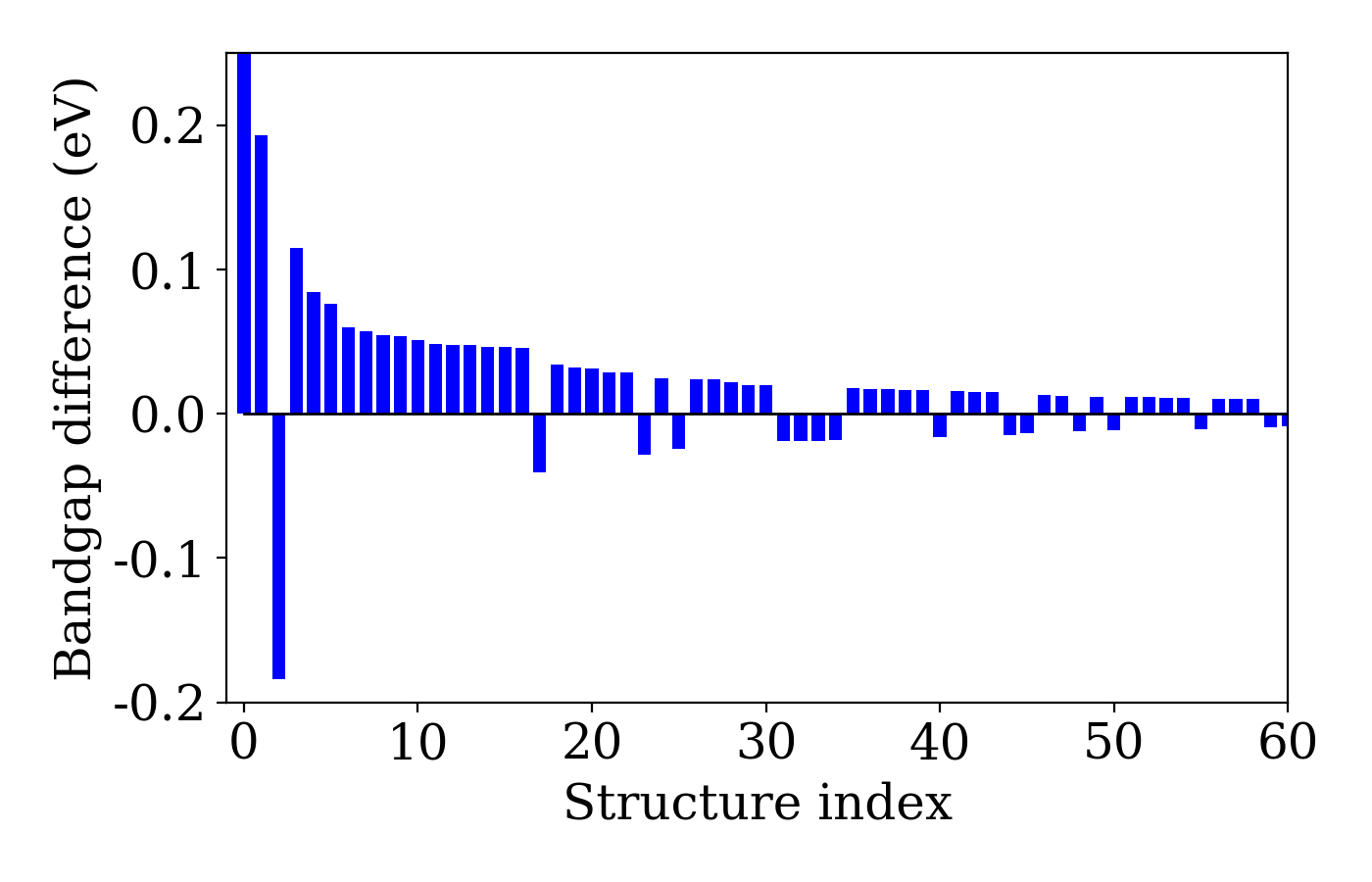} }
     \put( 0.2, 0.00){ \includegraphics[width=8.7cm, trim={0.0cm 0.5cm 0cm 0cm},clip=true]{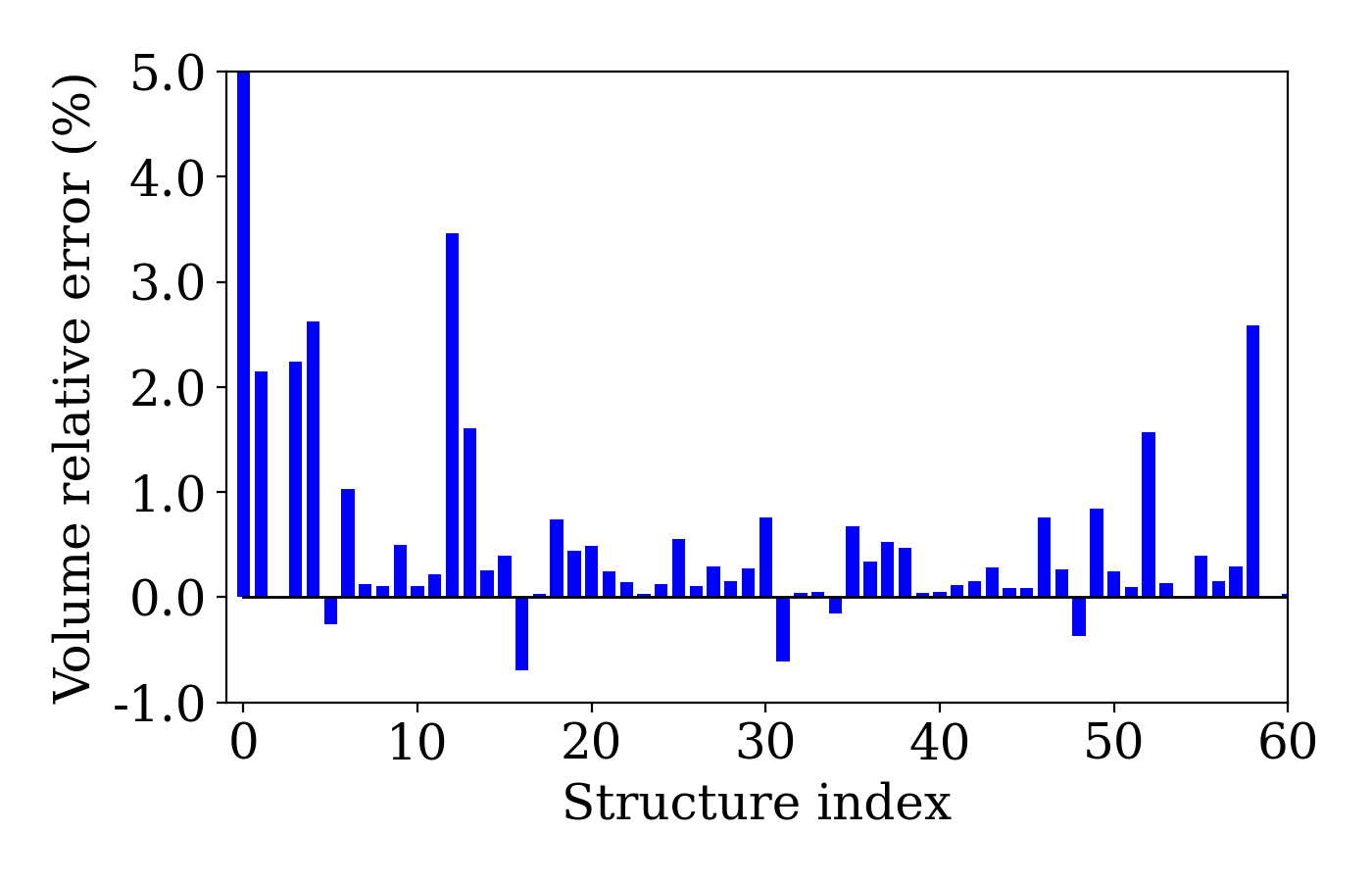} }
      \put( 2.5, 2.2){ \includegraphics[width=4.5cm, trim={0.0cm 0.5cm 0cm 0.5cm},clip=true]{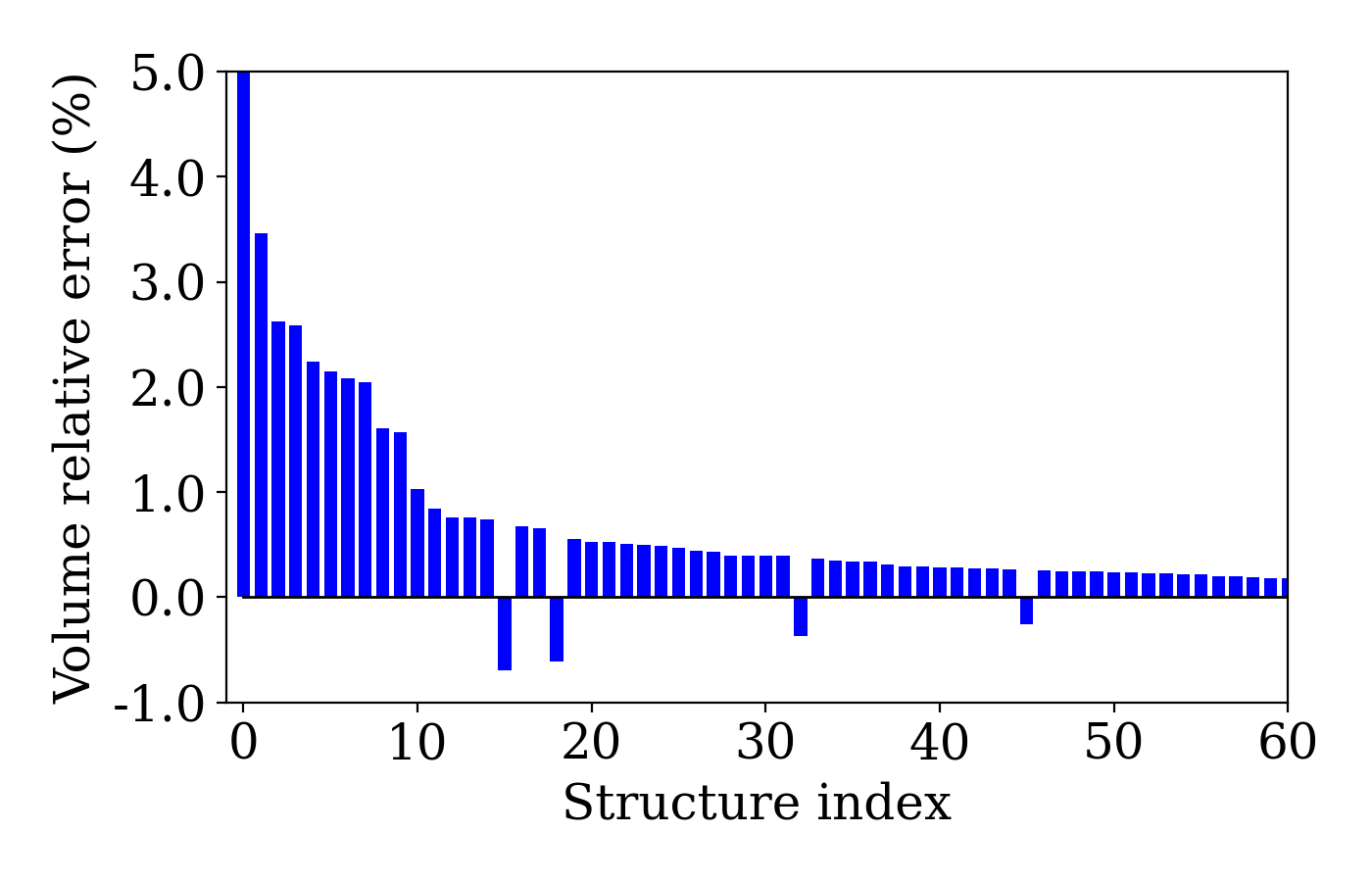} }
     \put(-0.1, 9.7){ \makebox(\boxsize,\boxsize){(a)} }
     \put(-0.1, 4.6){ \makebox(\boxsize,\boxsize){(b)} }
  \end{picture}
  \caption{\label{fig:difference_new_old_methods}
Comparison between the properties obtained with the bisection method and the new protocol to determine the Fermi energy in a random sample with 286 materials out of the $4,859 $ that had incorrect Fermi energies (out of the initial $24,842$).
(a) Bandgap difference obtained after full relaxation with the two methods.
The structures were sorted and indexed by the size of the difference.
A positive value indicates a new bandgap larger than the one obtained with the former protocol.
(b) The relative error in the unit cell volume with same structure indexing of (a).
In the inset, we sorted the structures by the volume relative error.
  } 
\end{figure}

We employed this new Fermi energy protocol to study some fundamental electronic properties in a database containing $24,842$ three-dimensions materials~\cite{huber_materials_2022}, including metals (54\%), insulators (46\%), and magnetic materials (21\%).
Technical details of the calculations are given in the next section.
The starting points are the relaxed structures fully converged using the former protocol, which is based on the bisection method only.
Then, we recalculated the systems' Fermi energy using the new protocol.
We obtained a different Fermi energy for 4859 materials (20\%), all of them semiconductors or insulators.
For the vast majority of materials in this group, we obtained atomic forces above the original relaxation threshold ($<10^{-5}$~eV/Å) with a median of 0.02 eV/Å and 10\% of the structures with forces higher than 0.13 eV/Å.

To elucidate the effect of the Fermi-energy correction on the properties of the materials, we performed a full relaxation using the new protocol for 286 randomly-selected materials (focus group) out of the 4859 identified above.
We focus on the variation of the bandgap and volume.
We chose to discuss the former, even at the Kohn-Sham DFT level, because the bandgap can readily signalize changes in the electronic structure due to the Fermi-energy modification.
In Fig.~\ref{fig:difference_new_old_methods}(a), we show the bandgap difference between the two approaches after full relaxation.
The new protocol produced both larger and smaller bandgaps in comparison with the former approach.
The median bandgap error was 0.002 eV while 10\% of structures resulted in an error of at least 0.022 eV.
This corresponds to a relative error in the bandgap of less than 1.78\% for 90\% of the structures.
In our small sample (286 structures), the two highest relative errors were for CGeI$_3$N, whose initial bandgap 1.16 eV increased by 80.14\%, and PbS$_2$, whose small bandgap 0.18 eV more than doubled.
We found no clear correlation between the relative error and the size of the initial bandgap.

Similarly, we also considered the relative error in the unit-cell volume, see Fig.~\ref{fig:difference_new_old_methods}(b).
The median relative error was 0.03\%, while 10\% of structures yielded an error larger than 0.43\%.
In this focus group, the highest volume errors were 25.10\% for CGeI$_3$N, corresponding to an absolute volume difference of 58.64~$Å^3$, and 3.46\% for CCs$_4$O$_4$, which is equivalent to an absolute difference of 12.48~$Å^3$.
As seen from Figs.~\ref{fig:difference_new_old_methods}(a) and (b), the volume relative error does not correlate with the error in the bandgap (the structures in both figures are indexed by the absolute value of the bandgap difference).

As cold smearing was used for these calculations, the undesirable Fermi energies obtained with the previous protocol are at or below the top of the valence bands.
This makes the very top of the valence under-occupied and just below over-occupied.
The corrected Fermi energies lie inside the band gap, which yields a uniform occupation of the states at the top of the valence bands thus affecting the self-consistent charge density.
It seems that in most cases, this Fermi energy correction leads to a volume expansion.
We could not identify a simple mechanism to predict when the correction would lead to an increase or reduction of the volume and bandgap.

\subsection{Details of the DFT calculations}
The high throughput calculation was managed with AiiDA~\cite{huber_aiida_2020,uhrin_workflows_2021}.
We used Quantum ESPRESSO v6.6~\cite{giannozzi_quantum_2009,giannozzi_advanced_2017} modified with the new protocol for Fermi energy determination.
We used the standard solid-state pseudopotentials family SSSP PBE Efficiency 1.1~\cite{prandini_precision_2018}.
The other parameters for the Quantum ESPRESSO simulation were determined by the protocol provided by the \emph{Quantum ESPRESSO input generator} available in the Materials Cloud~\cite{noauthor_quantum_nodate} and powered by AiiDA-QuantumESPRESSO plugin v3.0.0a3.
In particular, the kinetic energy cutoffs were the recommended values by the SSSP family.
K-point meshes with a minimum spacing distance of 0.15 1/Å were employed.
Cold smearing with degauss 0.01 Ry was used when not otherwise specified.
The data utilized and generated throughout this work is available in the Materials Cloud Archive~\cite{dossantos_Fermi_2023}.

\section{Conclusions}

In this work, we reviewed the fundamentals of the smearing technique used to improve the accuracy and robustness of DFT calculations.
We saw that advanced smearing methods such as Methfessel-Paxton and cold smearing, designed to reduce the coupling between total free energy and the smearing temperature, yield non-monotonic occupation functions.
As a consequence, the chemical potential can become non-uniquely defined, in particular in semiconductors and insulators at finite smearings.
Methfessel-Paxton smearing can lead to up to three distinct solutions, while cold smearing yields up to two.
Even if smearing is typically applied for metallic systems, automated or high-throughput calculations deal with materials where the band gap is unknown or can change during a self-consistent relaxation.
We demonstrated that algorithms to calculate the Fermi energy based on the bisection method could indeed identify an undesired spurious solution.
Hence, we proposed a numerical protocol employing Newton's minimization method to find the correct Fermi energy when using Methfessel-Paxton and cold smearing.
This protocol is now implemented in the open-source Quantum ESPRESSO package, and it is publicly available from release v6.8 onwards.
Finally, we conducted a high-throughput study with thousand of three-dimensional materials to validate the protocol presented here, highlighting how an incorrect Fermi energy can induce errors both in the relaxed crystal structure and its bandgap.
While in the majority of materials the error is minor, for a few cases, relative errors can be as large as 50\%.

\section{Acknowledgments}
F.J.d.S. acknowledges financial support from the European H2020 Intersect project through Grant No. 814487, and N.M. to the Swiss National Science Foundation (SNSF), through its National Centre of Competence in Research (NCCR) MARVEL.
We thank Marnik Bercx for helpful support on the high throughput study.

\bibliography{manuscript.bib}
\end{document}